\documentclass[twocolumn]{aa}
\usepackage{graphicx}

\usepackage{txfonts}
\begin{document}
\title{Testing Newtonian gravity with distant globular 
clusters: NGC1851 and NGC1904\thanks{Based on observations collected at the 
European Southern Observatory, Chile (Proposal 80.D-0106).}
}

   \subtitle{}

   \author{R. Scarpa\inst{1}\and G. Marconi\inst{2} \and G. Carraro\inst{2}
    \and R. Falomo\inst{3} \and S. Villanova\inst{4} }

\offprints{R. Scarpa; riccardo.scarpa@gtc.iac.es}

\institute{Instituto de Astrof\'isica de Canarias, Spain 
\and  European Southern Observatory, Chile 
\and Osservatorio Astonomico di Padova, Italy
\and Universidad de Concepcion, Departamento de Astronomia, Concepcion, Chile}

\date{\today}
\authorrunning{Scarpa et al.}
\titlerunning{Testing gravity in the weak acceleration regime}

\abstract{
  Globular clusters are useful to test the validity of
  Newtonian dynamics in the low acceleration regime typical of
  galaxies, without the complications of non-baryonic dark
  matter. Specifically, in absence of disturbing effects, e.g. tidal
  heating, the velocity dispersion of globular
  clusters is expected to vanish at large radii.  If such behavior is
  not observed, and in particular if, as observed in elliptical
  galaxies, the dispersion is found constant at large radii below a
  certain threshold acceleration, this might indicate a break down of
  Newtonian dynamics.
}{
  To minimize the effects of tidal heating that might increase the
  velocity dispersion at large radii, in this paper we study the
  velocity dispersion profile of two distant globular clusters, NGC
  1851 and NGC 1904.
}{
  The velocity dispersion profile is derived from accurate radial
  velocities measurements, obtained at the ESO 8m VLT telescope with
  the FLAMES multi-object spectrograph. Reliable data for 184 and 146 bona fide
  cluster star members, respectively for NGC 1851 and NGC 1904, were obtained.
}{
  These data allow to trace the velocity dispersion profile up to
  $\sim 2r_0$, where $r_0$ is the radius at which the cluster internal
  acceleration of gravity is $a_0\sim 10^{-8}$ cm s$^{-2}$. It is
  found that in both clusters the velocity dispersion is maximal at
  the center, decreases moving outward, and then becomes constant
  beyond $\sim r_0$.  Since the distance of these clusters from the
  Milky Way is large, the observed flattening of the velocity
  dispersion profile cannot be ascribed to tidal heating effects, as
  proposed in the case of nearer globular clusters.
}{
  These new results are fully in agreement with those found for other
  five globular clusters previously investigated as part of this
  project. Taken all together,  these 7 clusters support
  the claim that the velocity dispersion is constant beyond $r_0$,
  irrespectively of the specific physical properties of the clusters:
  mass, size, dynamical history, and distance from the Milky Way.  The
  strong similarly with the constant velocity dispersion observed in
  elliptical galaxies beyond $r_0$ is suggestive of a common origin
  for this phenomenon in the two class of objects, and might indicate
  a breakdown of Newtonian dynamics below $a_0$.
  \keywords{ Gravity --  Globular cluster -- star dynamics} 
}
\maketitle
%

\section{Introduction}
Stars within galaxies, and galaxies within clusters of galaxies are
very far apart from each other. As a consequence, the typical
gravitational accelerations governing their dynamics are orders of
magnitude smaller than the ones probed in our laboratories or in the
solar system. Thus, any time Newton's law of gravity is applied to
galaxies (e.g., to infer the existence of dark matter), its validity
is severely extrapolated. 

Although there are plenty of reasons to trust Newton's law in this
very weak acceleration regime, it is well known that spiral galaxies
rotation curves (see review by \cite{sofue01} and reference therein)
systematically deviate from prediction of Newtonian dynamics.
Similarly elliptical galaxies (e.g., \cite{Carollo95},
\cite{Mehlert00}, \cite{Pu10}) and cluster of galaxies velocity
dispersion profile (e.g., \cite{lokas06}), show a remarkable
flattening at large radii where a Keplerian falloff would be expected.
These deviations, ascribed to the existence of large amount of
non-baryonic dark matter (DM), appears to exhibit systematic (but not
yet understood) behaviors (c.f., recent findings by Gentile et
al. 2009 and by Donato et al. 2009). The most remarkable (e.g.,
\cite{binney04}) being that DM is needed to reconcile prediction with
observations when and only when the acceleration of gravity goes below
a {\it critical} value, of the order of $a_0 \sim 10^{-8}$ cm s$^{-2}$
(\cite{begeman91}).

This systematic suggests we may be facing a breakdown of Newton's law
rather than the effects of DM. If this is the case, one should observe
the same phenomenology in all the systems where the acceleration goes
below $a_0$.  In particular, the detection of a systematic flattening
of the velocity dispersion profile in the external region of globular
clusters, which are known to contain negligible quantities of DM,
would represent a strong indication of a failure of the adopted law
rather than the effects of unseen matter.

In 2003 Scarpa, Marconi and Gilmozzi (2003A,B) presented the results
of a pilot study of the dynamics of the external region of the
globular cluster $\omega$ Cen.  The velocity dispersion was traced up
to 45 pc from the center, probing acceleration as small as $\sim
7\times 10^{-9}$ cm s$^{-2}$. Clear evidences were found that, as soon
as the internal acceleration of gravity approached $a_0$, the velocity
dispersion did not vanish, converging toward a constant value.  A
claim that was recently disputed by Sollima et al. (2009), but
reconfirmed by Scarpa and Falomo (2010) by a reanalysis of all
available data.

The behaviors seen in $\omega$ Cen is not unique and appears to occur
in all globular clusters for which an adequate analysis of the
velocity dispersion profile was done: NGC 6171 and NGC 7078 (Scarpa,
Marconi \& Gilmozzi 2004A,B), NGC 7099 and NGC 6341 (\cite{Scarpa07a},B).

Even more intriguing is the case of the low concentration cluster
NGC288. This cluster has internal accelerations of gravity every where
below $a_0$ and thus is a tiny version of a low surface brightness
spheroidal galaxy.  Similarly to these galaxies (\cite{Mateo97};
\cite{Wilkinson06}; \cite{Koch07}) NGC 288 has within errors the same
velocity dispersion at all radii (\cite{Scarpa07b}).  Another
low-density cluster possibly allowing deviations from 
Newtonian dynamics is Palomar 14 (\cite{Gentile09}).

These results point out an intriguing correspondence between
elliptical galaxies and globular clusters.  Dense globular clusters do
probe the same accelerations and may behave like high surface
brightness galaxies, while loose clusters do probe the same
accelerations and may behave like low surface brightness
galaxies. What is different is the interpretation of the data. The
common wisdom goes that in galaxies it is the dark matter that alters
the dynamics, while in globular clusters something else does it. 

The most viable explanation being tidal heating, the
  increase of the stellar velocity dispersion due to the differential
  acceleration produced by the Milky Way in different position of the
  globular cluster.  It would be, however, more logic to invoke a common
origin of the phenomenon, that is a failure of Newtonian dynamics
below $a_0$ as claimed within the frame work of MOND
(\cite{Milgrom83}).

In this paper, to further generalize our investigation, we present new
accurate radial velocity measurements for NGC 1851 and NGC 1904. These
clusters were selected for being located at r=16.7 and 18.8 kpc from
the galactic center, respectively.  Compared to the other globular
clusters studied so far, NGC 1851 and NGC 1904 are approximately twice
as distant from the Milky Way center. Thus are experiencing a tidal
heating, proportional to $r^{-3}$, about one order of magnitude
smaller, making its effects negligible.

\begin{figure}
\centering \includegraphics[height=7cm]{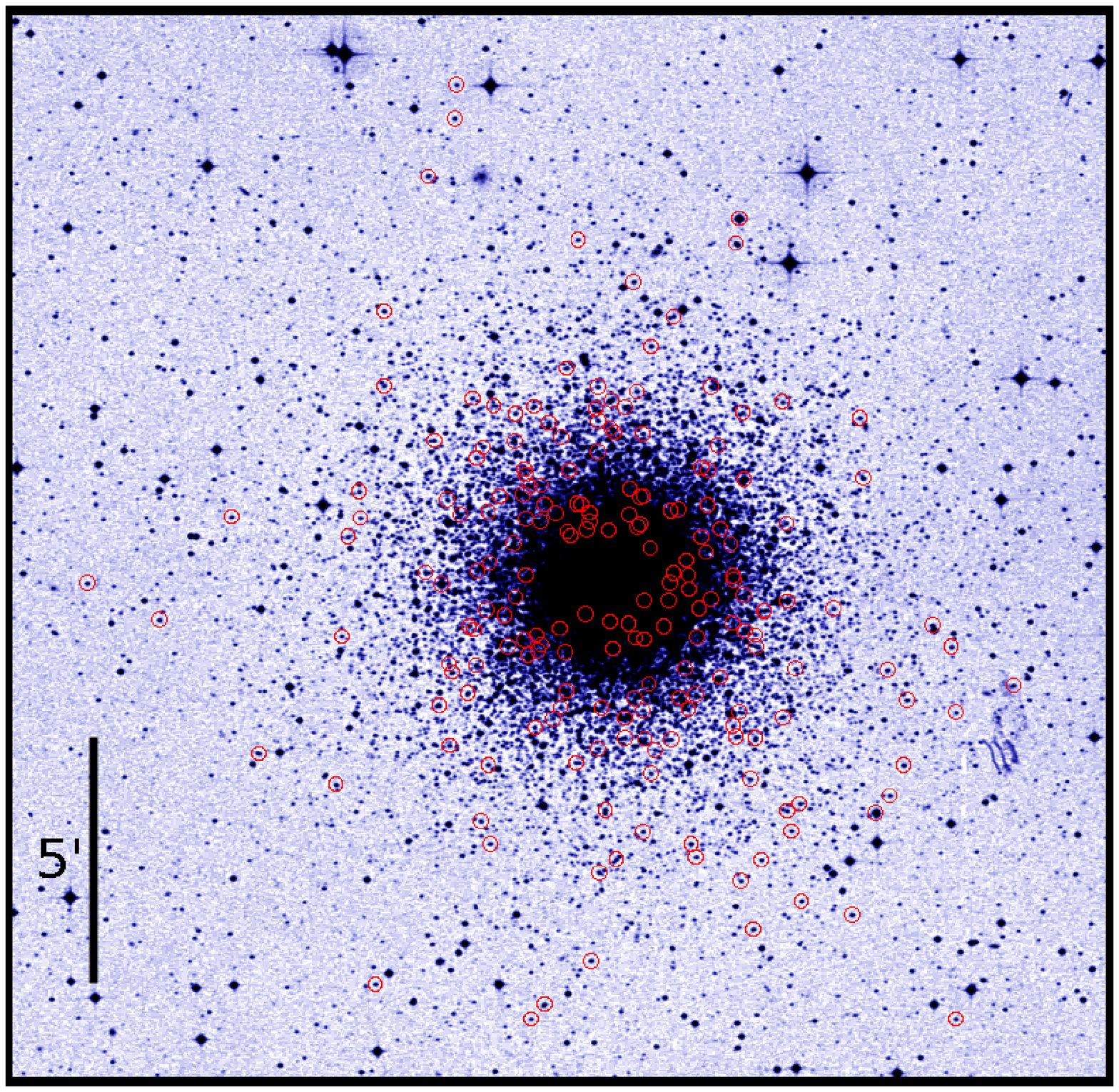}
\includegraphics[height=8.5cm]{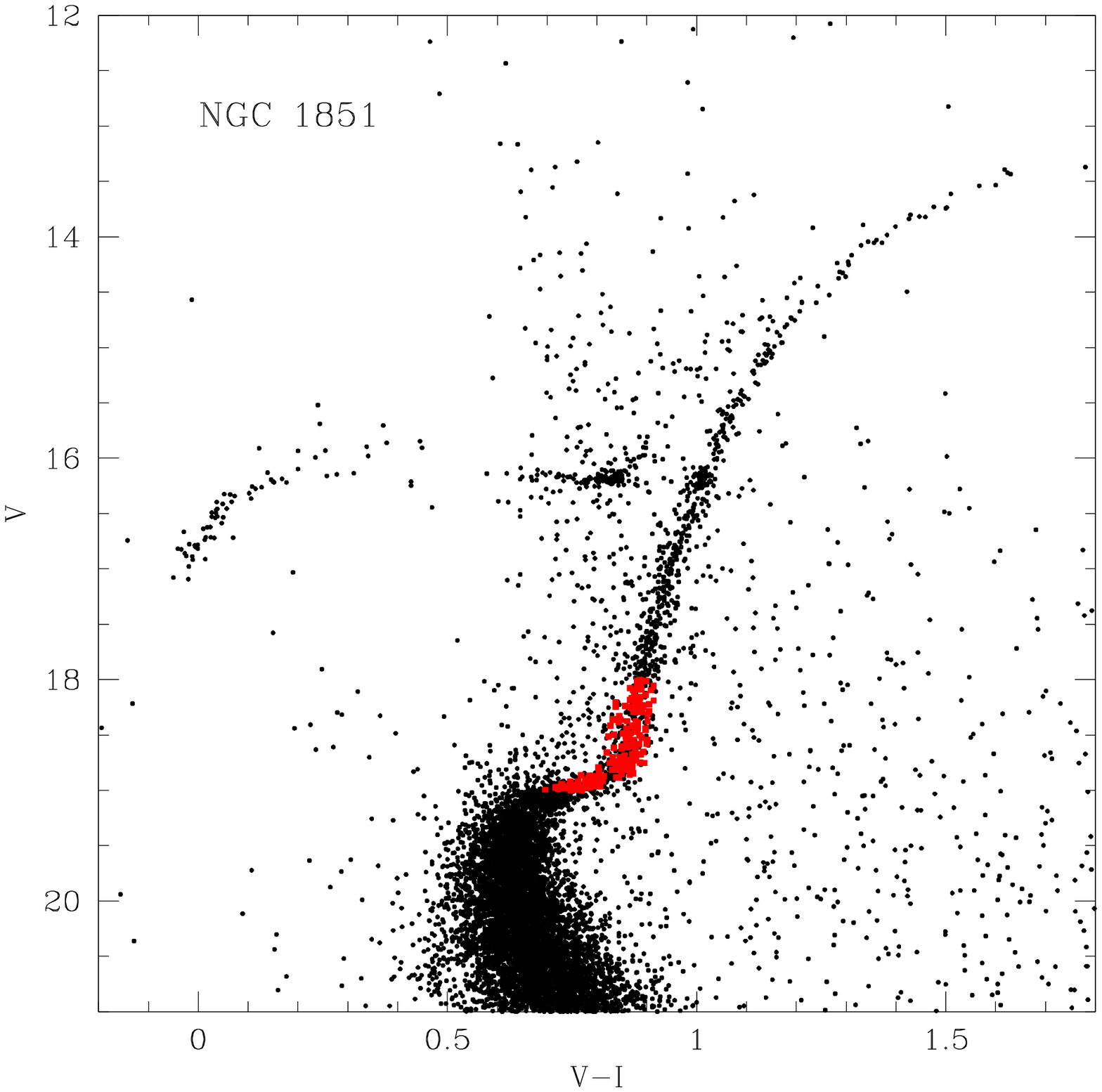}
\caption{\label{ngc1851_HR} The distribution of the selected stars in
  the cluster NGC1851 (north on the top, east on the left), and the
  color magnitude diagram with the selected targets highlighted in
  red. For reference, the radius where the acceleration of gravity is
  $a_0$ is 2.9 arcmin.}
\end{figure}

\begin{figure}
\centering \includegraphics[height=7cm]{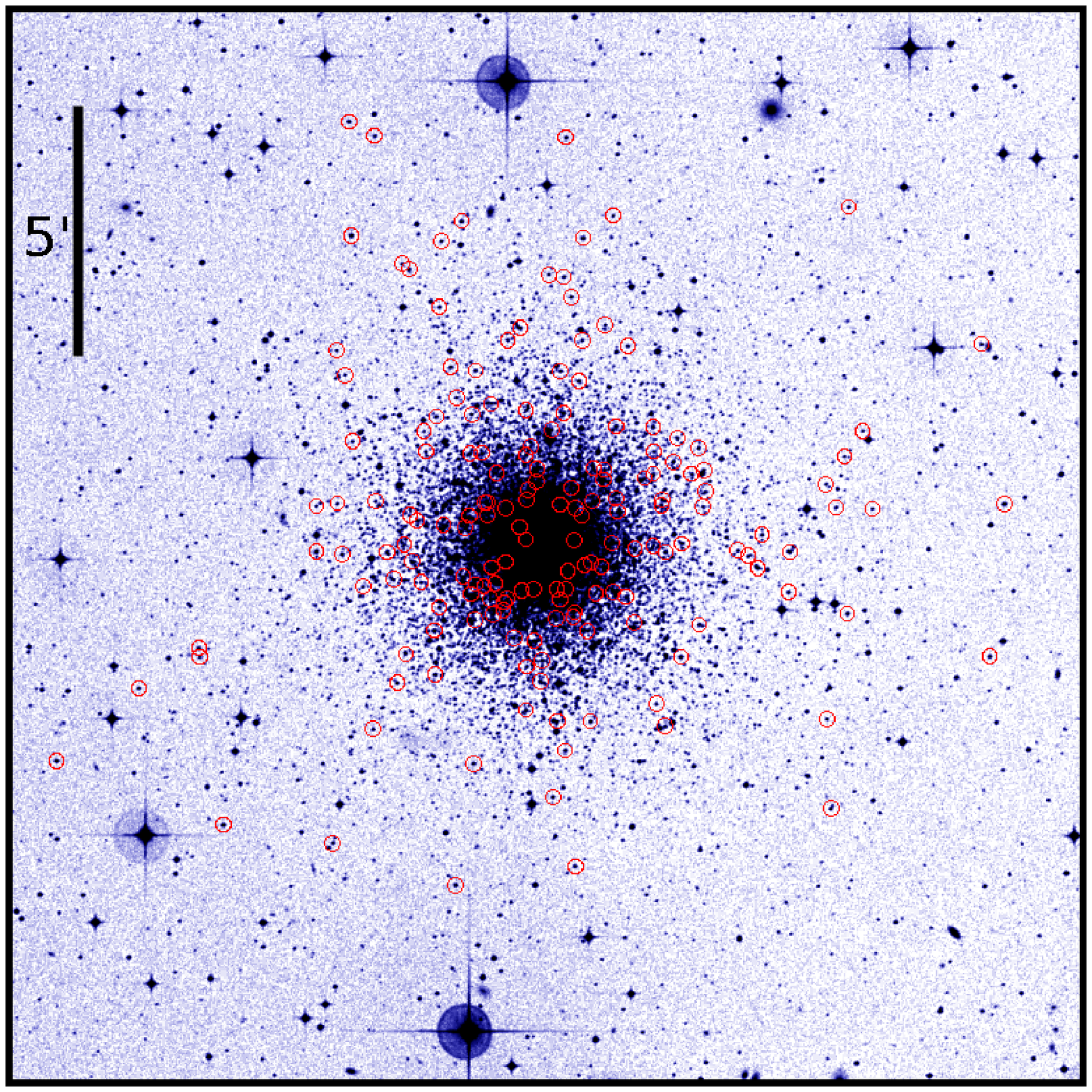}
\includegraphics[height=8.5cm]{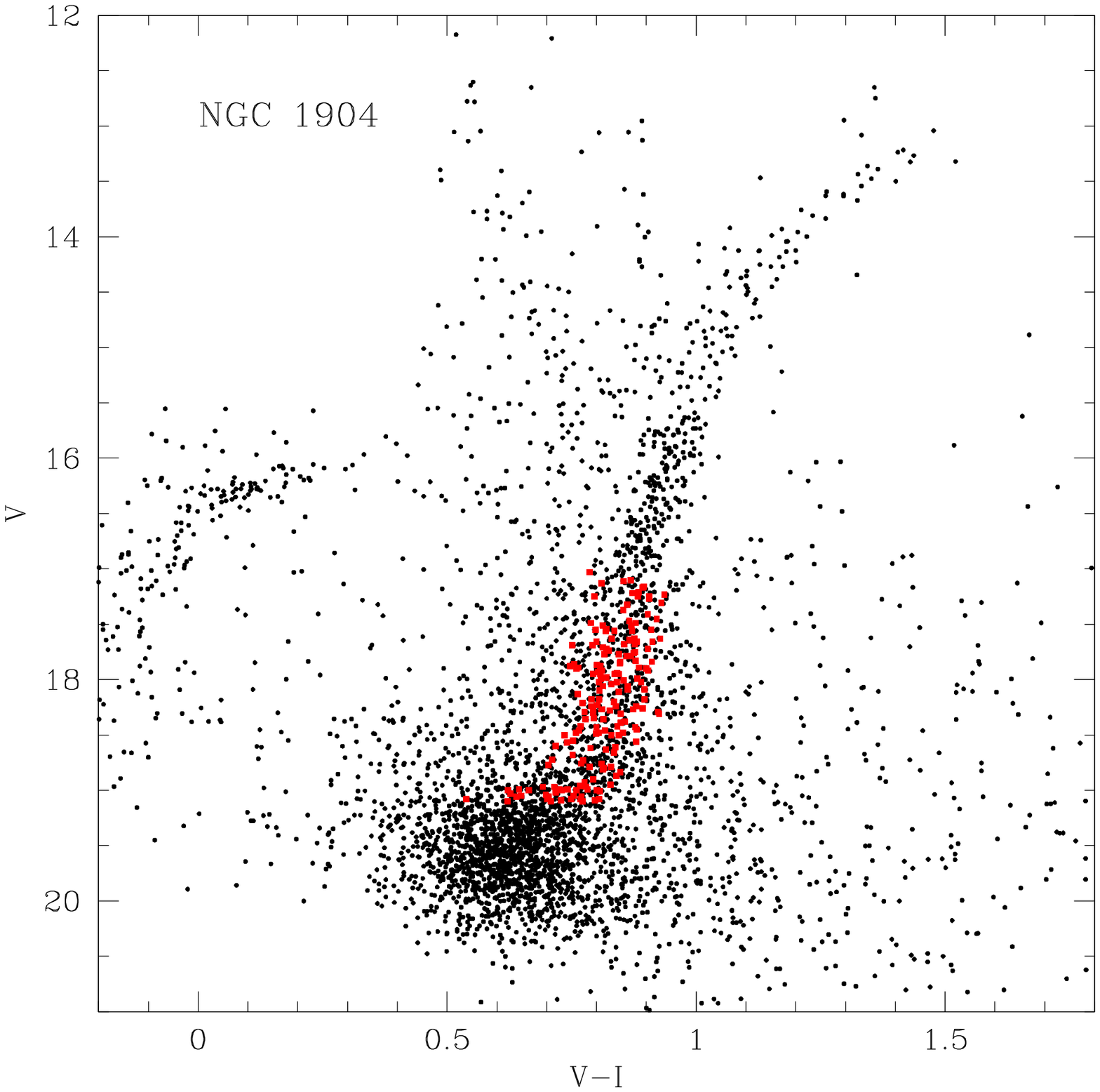}
\caption{\label{ngc1904_HR} 
The distribution of the selected stars in the cluster NGC1904 
(north on the top, east on the left), and
the color magnitude diagram with the selected 
targets highlighted in red. For reference, the radius where 
the internal acceleration of gravity is  $a_0$ is 2.1 arcmin.}
\end{figure}

\section{Observation  and data analysis}

The initial selection of targets was based on their color, as derived
from the analysis of ESO Imaging Survey frames and ESO 2.2m Wide Field
Imager data.  Targets have been selected requiring color difference
from the cluster main sequence $V-I<0.05$ and $V-I<0.1$, and apparent
magnitude of $19>m>18$ and $19>m>17$, respectively for NGC 1851 and
1904.  The cut in luminosity was made close to the base of the giant
branch to probe the cluster stellar population in a well populated
region to ensure good probability to find cluster members at large
distances from the cluster. Indeed, according to Milky Way stellar
population models \cite{Vanhollebeke09} we expect a contamination of
only 0.029 and 0.118 stars per arcmin squared in the selected
color-luminosity range. With this surface density we expect a
contamination of about 1.8\% and 7.5\% of our initial target list,
respectively for NGC 1851 and 1904.

In Figures \ref{ngc1851_HR} and \ref{ngc1904_HR} we show the color
magnitude diagram (CMD) of the clusters with highlighted the position
of the 199 and 173 stars that were actually observed, respectively in
NGC 1851 and 1904. The final distribution of the targets around the
cluster center is also shown.

Because of the moderately high galactic latitude and the good
photometric data we did expect little contamination and very high
success rate in identifying cluster members.

Spectroscopic observations were then obtained with FLAMES+GIRAFFE
(\cite{pasquini02}) at the ESO 8 meters VLT telescope. FLAMES is a fiber
multi-objects spectrograph, allowing the simultaneous observation of
up to 130 objects. We selected the HR9B setup that includes the
magnesium triplet covering the wavelength range $5143 < \lambda <
5346$ at resolution R=25900. Stellar astrometry was derived from the
US Naval Observatory (USNO) catalog, which proved to have the required
accuracy (0.3 arcsec) for FLAMES observations.  Two different fiber
configurations were necessary to allocate all the selected stars. For
each configuration 5 separate 3200 s exposures were obtained in
different night from November 2007 to March 2008, under good
atmospheric condition (clear sky, airmass $<1.5$, and seeing $\sim 1$
arcsec).

Data were automatically pre-reduced using the GIRAFFE pipeline
GIRBLDRS (Blecha et al. 2000; see http://girbldrs.sourceforge.net for
GIRAFFE pipeline, software and documentation), in which the spectra
have been de-biased, flat-field corrected, extracted, and
wavelength-calibrated, using both prior and simultaneous
calibration-lamp spectra. A sky correction was applied to each stellar
spectrum by subtracting the average sky spectra obtained through
dedicated fibers.  The resulting spectra have dispersion of 0.05 
\AA /pixel.

Radial velocities were obtained from the IRAF fxcor cross-correlation
task. Stellar spectra were cross-correlated with a synthetic
template calculated by SPECTRUM, the LTE spectral synthesis program
freely distributed by Richard O. Gray (Program and documentation
available at www.phys.appstate.edu/spectrum/spectrum.html), for the
mean temperature, gravity $\log{g}$, and metallicity of our stars. We
verified on the Kurucz solar flux atlas and UVES solar spectra 
that the template was
accurate at the level of 50 m/s or less. The accuracy of the data
reduction procedure and of the radial velocity measurement was
extensively tested in previous works (e.g. \cite{Milone06},
and \cite{Sommariva09}).

\begin{figure}
\centering
\includegraphics[height=11cm]{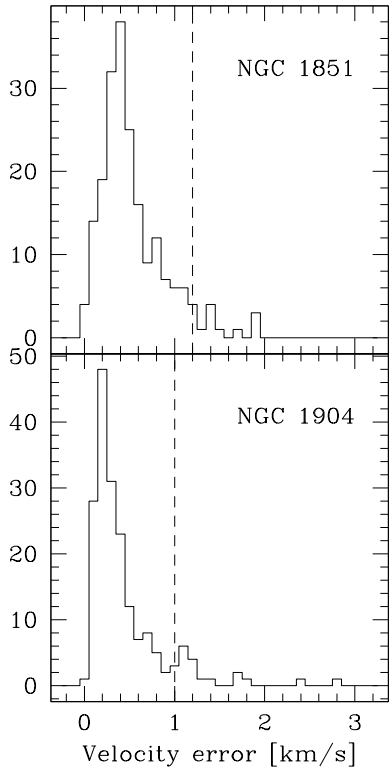}
\caption{\label{figErrors} Distribution of velocity uncertainties as
  derived from the comparison of repeated velocities measurements of
  the same target.  The peak of the distribution is at $\sim 0.4$ and
  $\sim 0.2$ km/s in NGC 1841 and 1904, respectively. This difference
  is due to targets in NGC 1904 to be on average 0.4 magnitudes
  brighter than those in NGC 1851.  The dashed vertical line is the
  adopted threshold for rejecting objects.}
\end{figure}

Each spectrum was treated independently from the others, so that we
ended up with 5 independent measurement of the radial velocity for
each star (see Appendix). For a limited number of stars we have less
than 5 spectra because some were rejected for being affected by strong
cosmic rays and other defects. The exact number of spectra available
for each star is given in tab \ref{tabVel1851} and
\ref{tabVel1904}. These radial velocities were then used to quantify
the effects of possible intrinsic variation of the radial velocity
(e.g., due to binary stars, stellar photospheric winds, measurement
errors, etc.) finding for the vast majority of the targets a
dispersion of less than 2.0 and 1.5 km/s in the case of NGC 1851 and
NGC 1904, respectively (Fig. \ref{figErrors}). This correspond to a
statistical uncertainty on less than 1 km/s on the average radial
velocity of each target, small enough for not affecting the
measurement of the velocity dispersion of the cluster even in the most
external regions.  In the following, we define as best estimate of the
true radial velocity the weighted average of the 5 independent
measurements.

A number of stars (5 in NGC1851 and 8 in NGC1904) were included in
both FLAMES fibers setup to double check whether a velocity shift
could exists between the two data sets. No statistically significant
shift was detected down to 100 m/s.  All velocities 
presented here are heliocentric.

\section{Results}

To derive the velocity dispersion profile (VDP) we first separated
members from non-members in the velocity space.  There were no
ambiguity since the distribution of radial velocities was found
bimodal, with separation of $\sim 100$ km/s. In total, 4 non-members
(2\% of the sample) where found in NGC 1851 and 14 (8\% of the sample)
in NGC 1904, confirming a posteriori the high efficiency of our color
selection criteria and in agreement with expectation from galaxy
stellar population models (see section 2).

The median of the radial velocity error distribution for all members
is 0.43 and 0.29 km/s (Fig. \ref{figErrors}), significantly smaller
then the expected velocity dispersion we are trying to measure (see
next section). The radial velocity error for a small fraction of the
targets deviate significantly from the rest of the distribution. Thus
we decided to retain only measurements with error smaller than four
times the semi-inter quartile range of the error distribution (i.e, 1.2
and 1.0 km/s for NGC 1851 and 1904).

\subsection{NGC 1851}

\begin{figure}
\centering
\includegraphics[height=6.5cm]{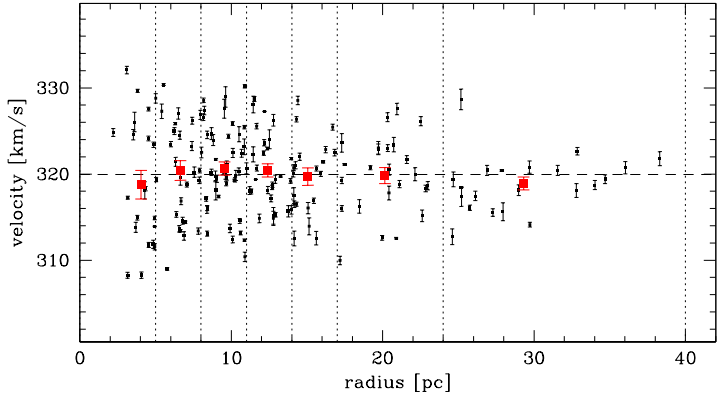}
\caption{\label{velocity1851} The distribution of the 184 targets with
  radial-velocity uncertainty $<1.2$ km/s as a function of distance
  from the center of NGC 1851 (Points with error bar). The large squares with
  error bar represent the average velocity in the bins indicated by
  vertical lines.  The horizontal dashed line represents the cluster mean
  radial velocity. }
\end{figure}

From the initial sample of 199 stars, after eliminating 4 non-members
cutting on radial velocity, and retaining only stars with velocity
accuracy better than 1.2 km/s , we are left with 184 reliable radial
velocities that represent the final sample used to constrain the
cluster dynamics. The cluster average radial velocity is $320.0\pm
0.4$ km/s, with r.m.s. of 4.9 km/s, in good agreement with the value
$320.5 \pm 0.6$ km/s (\cite{harris96}). The average radial-velocity within
each bin shows no systematic trends as a function of distance from the
cluster center (Fig. \ref{velocity1851}).

\begin{figure}
\centering
\includegraphics[height=9cm]{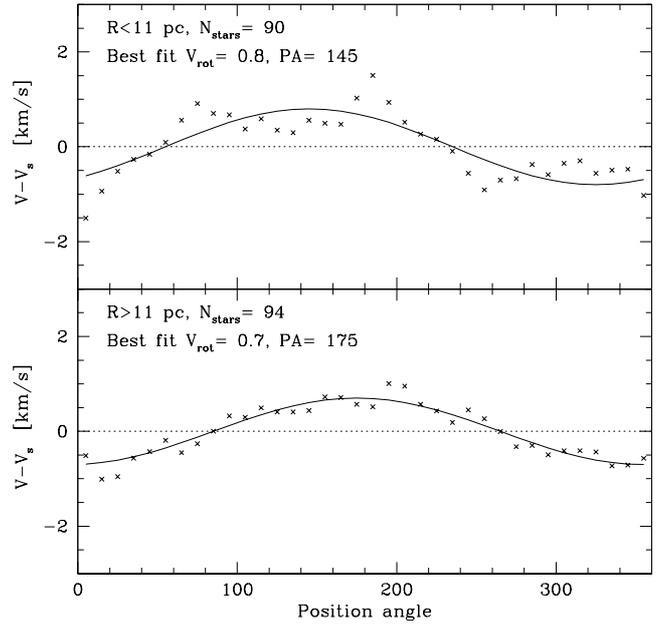}
\caption{\label{rot1851} Rotation in the inner and outer regions of
  NGC1851.  The plotted value is the average radial velocity of stars
  in a 180 degree sector, minus the average velocity of the whole
  cluster. The half cluster sector has been moved by 10 degrees from
  one point to the next.  The maximum rotational velocity in km/s is
  derived from the best sinusoidal fit of the data. The position angle
  (in degrees from North toward East) of the rotation axis is also
  shown.}
\end{figure}

The cluster is slowly rotating, with a maximum ordered rotation
velocity of less than $0.8$ km/s, both in the inner and external
regions (Fig. \ref{rot1851}). Such a small rotational velocity cannot
contribute significantly to sustain the structure of the cluster and
will be therefore neglected in the following discussion.

\begin{figure}
\centering
\includegraphics[height=6cm]{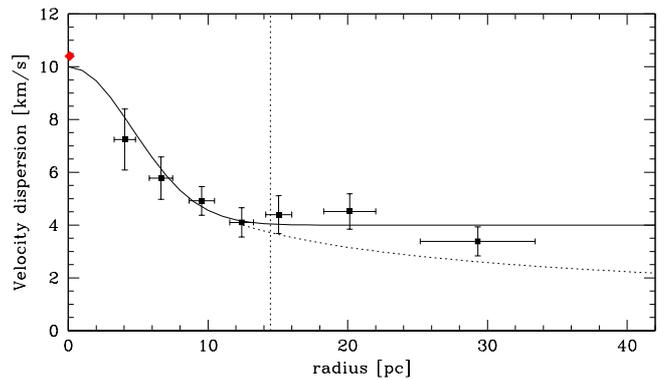}
\caption{\label{disp1851}
  NGC 1851 velocity dispersion computed binning the 
  data as in Fig. \ref{velocity1851}. 
  Error bars along the X axis represent the dispersion of the
  data within the bin, along the Y axis they give the 1-$\sigma$
  uncertainty on the dispersion.  The central velocity dispersion as
  given in \cite{harris96} is also shown (dimond).  The vertical line marks the
  radius where the acceleration is $a_0$. Over plotted to the data is a
  Gaussian plus a constant, which is not a fit to the data, meant to
  better highlight the flattening of the velocity dispersion
  that occur well within the cluster tidal radius of 41pc (\cite{harris96}).
  }
\end{figure}

\begin{figure}
\centering
\includegraphics[height=6cm]{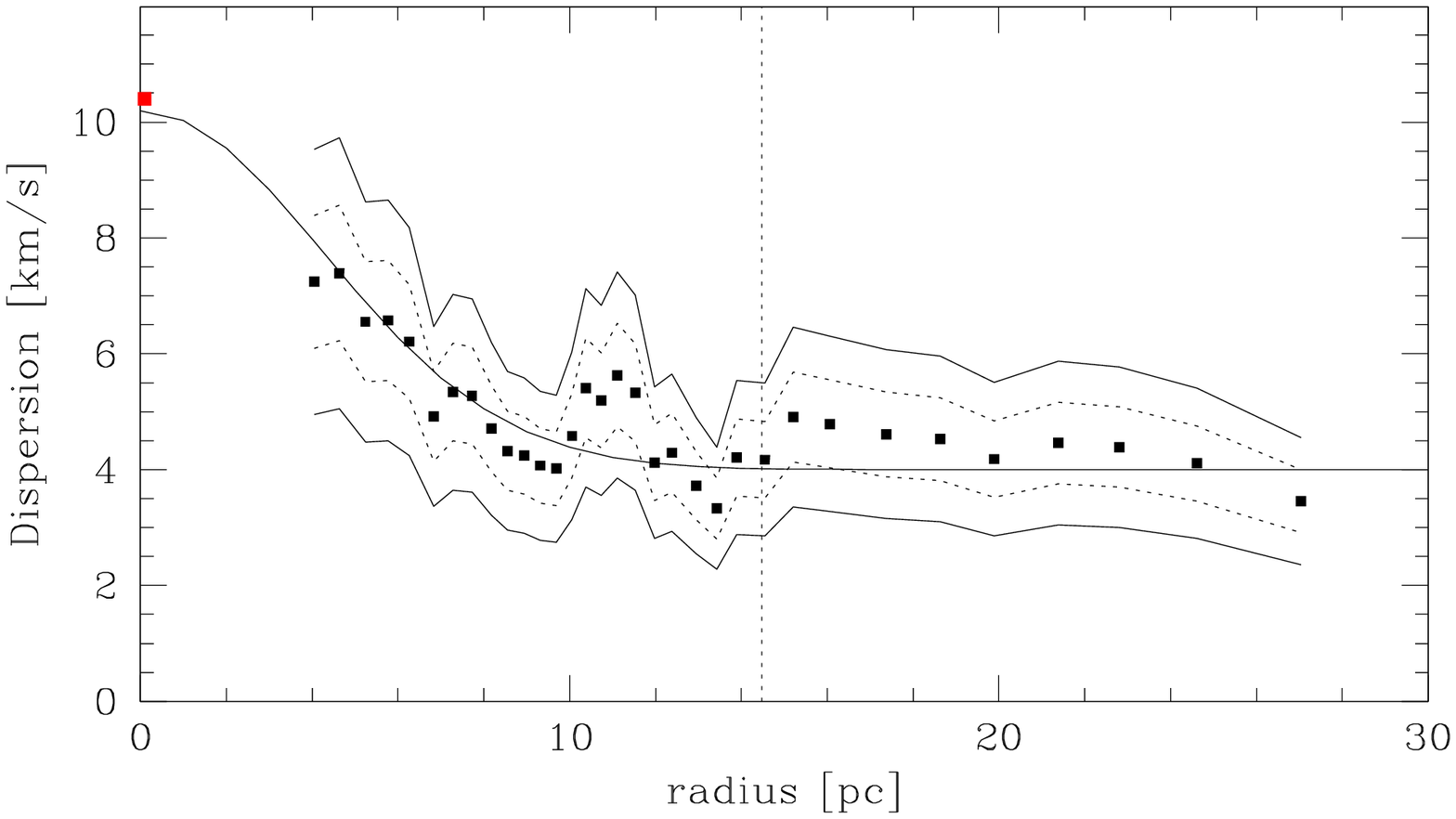}
\caption{\label{runningDisp1851} The velocity dispersion profile of
  NGC1851, derived considering the 184 members with radial velocity
  accuracy better than 1.2 km/s.  Data have been binned in groups of
  20 each, moving by 5 from one group to the next.  The dotted and
  solid lines give the 66 and 90 percent confidence regions for the
  dispersion. The central velocity dispersion as reported in
  \cite{harris96} is also shown (dimond). The vertical line marks the
  radius where the internal acceleration of gravity is $a_0$ (assuming
  M/L=1).  Over plotted to the data is a Gaussian plus a constant (not
  a fit to the data) used to better highlight the flattening of the
  VDP that occur well within the cluster tidal radius of 41pc (\cite{harris96}) . }
\end{figure}

The velocity dispersion was computed in a number of selected bins
following the procedure described in (Scarpa and Falomo 2010).  Since
the shape of the VDP might depend on the choice of the binning, we
investigated the effects of different binning on the VDP.  Two
representative cases are shown.  The first (Fig. \ref{disp1851}) was
obtained grouping the data using integer numbers in radial
distance. The binning limits, the number of objects in each bin, and
the corresponding velocity dispersion are given in table
\ref{tabDisp1851}.  In the second case (Fig. \ref{runningDisp1851}),
data have been grouped assigning a fix number of object to each bin,
20 in this case, and moving by 5 objects from one bin to the next.
This ensure a much higher sampling of the profile, because data are
partially reused moving from one bin to the other, at the expenses of
having statistically correlated values. While it is clear that due to
the limited number of data the velocity dispersion fluctuates quite
significantly, the general trend is well defined. Toward the center
the velocity dispersion naturally extrapolates to the central velocity
dispersion available in the literature (10.4 km/s,
\cite{harris96}). In the external regions, the dispersions flattens
out converging toward a constant value of $\sim 4.0 \pm 0.5$ km/s.  Moving from
the center outward, the VDP exhibits fluctuations of up to $\sim 2$
km/s, making difficult to pin down the radius where the flattening
first occur. This can be as close as 10 arcmin from the center, or as
far as 15 arcmin.  Assuming M/L=1 in solar units, total V-band
absolute magnitude for the cluster of M=-8.33, and distance from the
sun 12.1 kpc (\cite{harris96}), the internal acceleration of gravity
is $a_0$ at r=14.5 pc from the center. Thus, data are consistent with
a flattening of the dispersion profile where $a<a_0$.

\subsection{NGC 1904}
\begin{figure}
\centering
\includegraphics[height=6.5cm]{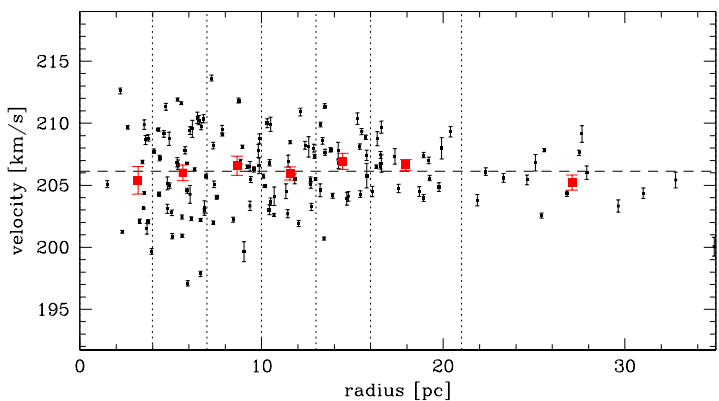}
\caption{\label{velocity1904} The radial velocity distribution of the 146
  targets with radial-velocity accuracy better than 1.0 km/s (dots) as a
  function of distance in parsecs from the center of NGC 1904.  The
  squares with error bar represent the average velocity in each bin
  indicated by the vertical lines.  The horizontal dashed line gives
  the cluster mean radial velocity.}
\end{figure}

From the initial sample of 173 stars, the cut on radial velocity and
velocity accuracy better than 1.0 km/s left us with 146 targets used
to study the cluster dynamics. The average radial velocity of NGC 1904
is $206.1\pm 0.3$ km, with r.m.s. of 3.1 km/s, in good agreement with
the value available in the literature ($206 \pm 0.4$,
\cite{harris96}).  No systematic trend are seen in the average
radial-velocity distribution as a function of distance from the
cluster center (Fig.  \ref{velocity1904}).

\begin{figure}
\centering
\includegraphics[height=9cm]{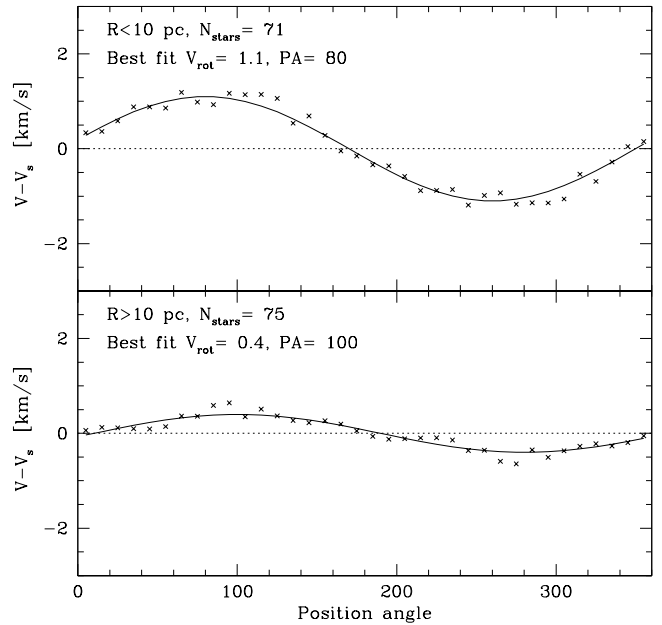}
\caption{\label{rot1904} Rotation in the inner and outer regions of
  NGC1904.  The plotted values represent the difference between the
  average radial velocity of stars in a 180 degree sector and the
  systemic velocity of the whole cluster.
  The maximum rotational velocity in km/s is
  derived from the best sinusoidal fit of the data. The position angle
  (in degrees from North toward East) of the rotation axis is also
  shown.}
\end{figure}

The cluster is slowly rotating (Fig. \ref{rot1904}). Within 3 arcmin from
the center we measured a rotation velocity of 1.1 km/s, with position
angle 85 degrees (from North toward East). Outward of this the best
sinusoidal fit is consistent with no rotation at all.  We can then
neglect the rotation in the outer part of the cluster, the most
relevant for this study.

\begin{figure}
\centering
\includegraphics[height=6cm]{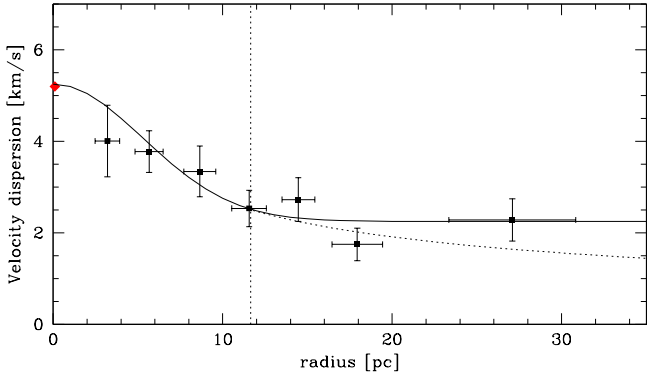}
\caption{\label{disp1904} The velocity dispersion of NGC 1904 computed
  with the same binning used in Fig. \ref{velocity1904}. Error bars
  along the x axis represent the dispersion of the data within each
  bin, along the y axis they give the 1-sigma uncertainty on the
  dispersion.  The central velocity dispersion as given in
  \cite{harris96} is also shown (dimond).  The vertical dotted line marks the
  radius where the acceleration is $a_0$. Over plotted to the data is
  a Gaussian plus a constant, not a fit to the data, meant to better
  highlight the flattening of the velocity dispersion
  that occur well within the cluster tidal radius of 31pc (\cite{harris96}). }
\end{figure}

\begin{figure}
\centering
\includegraphics[height=6cm]{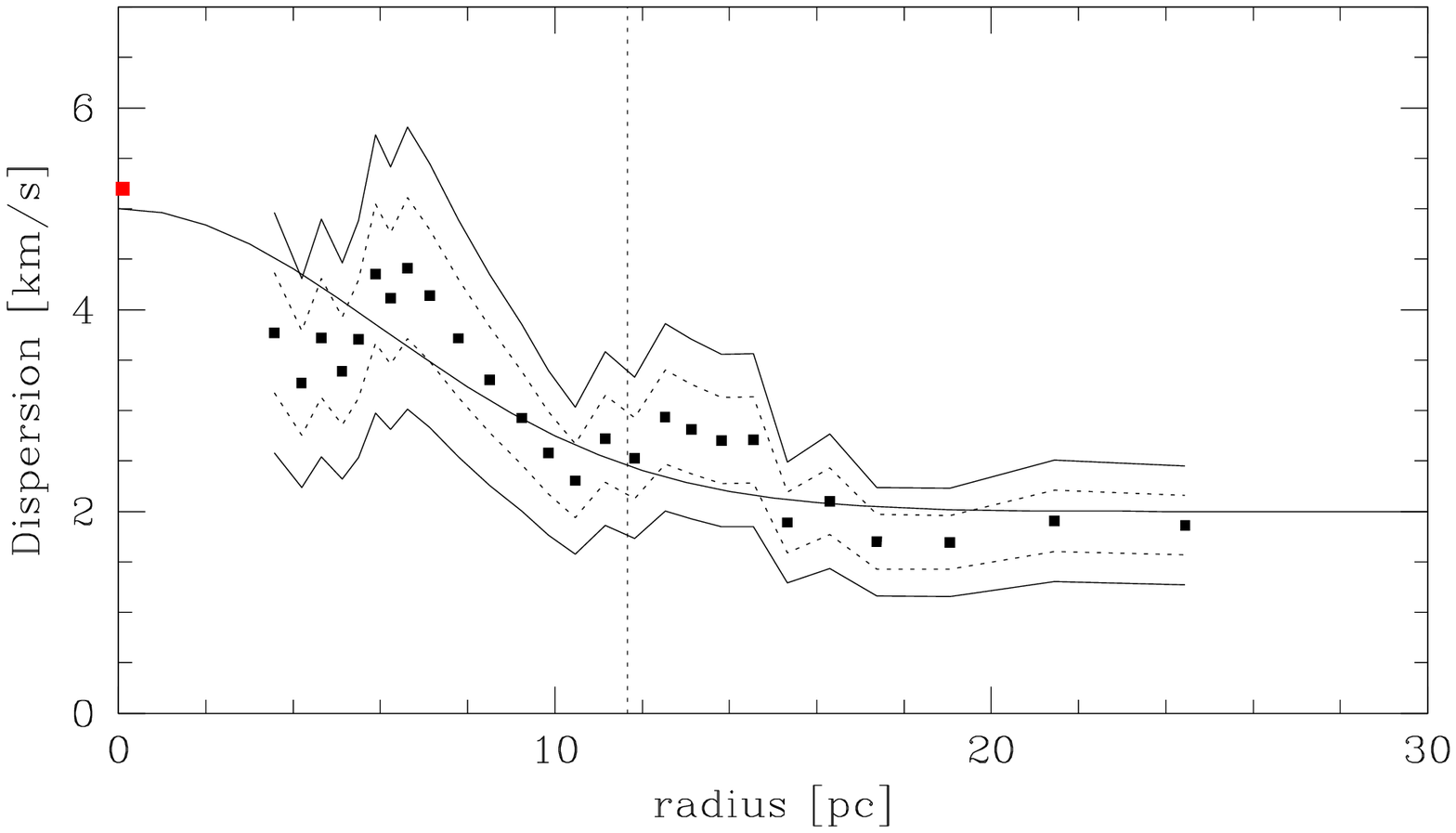}
\caption{\label{runningdisp1904} Velocity dispersion profile of
  NGC1904 as derived binning data in groups of 20 each, moving by 5
  from one group to the next.  The dotted and solid lines give the 66
  and 90 percent confidence regions for the dispersion. The central
  velocity dispersion as reported in \cite{harris96} is also
  shown (dimond). The vertical line indicates the radius where the 
  internal acceleration of
  gravity is $a_0$ (assuming M/L=1).  Over plotted to the data is a
  Gaussian plus a constant, not a fit to the data, used to better
  highlight the flattening of the VDP 
  that occur well within the cluster tidal radius of 31pc (\cite{harris96}).}
\end{figure}

As for NGC1851 we present the VDP derived with two different
binning approach (Fig. \ref{disp1904} and \ref{runningdisp1904}).  Due
to the low mass of NGC1904 cluster, the dispersion is quite small and
it is difficult with the present data set to derive a compelling
result. What is seen in all binning approaches used, is that toward
the center the profile might be consistent with central velocity
dispersion of 5.2 km/s reported in the literature (\cite{harris96}),
but clearly due to the large fluctuations it is difficult to draw any
firm conclusion.  Moving outward, the dispersion goes up and down a
couple of times before converging toward a constant value of $2.25 \pm 0.3$
km/s.  Basically, starting from 10 pc from the center the dispersion
does not change much, ranging within $1.8 < \sigma < 2.6$ km/s.

Assuming M/L=1 in solar units, total V-band absolute magnitude for the
cluster of M = $-7.86$, and distance from the sun 12.9 kpc
(\cite{harris96}), the internal acceleration of gravity is $a_0$ at
r=11.6 pc from the center. Thus, also in this cluster data suggest a
flattening of the VDP might occur at this particular value of the
acceleration.

\section{Discussion}

We have reported radial velocity measurements for two moderately
distant globular clusters, NGC 1851 and 1904, located at 16.7 and 18.8
kpc from the galactic center, respectively.  The analysis of these
data indicates that in these clusters the velocity dispersion remains
constant beyond $r=12.5\pm 2.5$ and $r=12\pm 2$ pc from the center, 
respectively. Values very similar and statistically fully consistent
with the radius $r_0$ where the acceleration is $a_0$.

Moreover, the two clusters are moving in a fast receding orbit.
Dinescu et al (1999) assuming a logarithmic gravitational potential
for the Milky Way, computed the most probable orbit for NGC 1851 and
1904. Integration of these orbits shows that they passed
perigalacticon about 51 and 86 millions years ago, respectively.
According to our velocity dispersion measurement, even in the
outermost regions of the clusters the star velocity dispersion is
$3.4$ and $2.3$ km/s.  With this velocity stars in NGC 1851 and 1904
cover a distance of twice the tidal radius in $\sim 24$ and $\sim 32$
Myr, shorter than the time since last perigalacticon. Hence these
clusters had enough time to revirialise.  Moreover, these two clusters
experience a tidal action due to the the Milky Way about one order of
magnitude smaller than that acting on the clusters previously studied
as part of this project, making it unlikely that the flattening of the
VDP is the result of tidal heating.

Thus, these new observations bring to 7 out of 7 the number of high
concentration globular clusters showing constant velocity dispersion
at large radii (Table 6).

While differing in many respects (mass, dynamical history,
concentration, position in the Milky Way halo, etc.), these clusters
do share the property of being sufficiently concentrated to have
internal accelerations of gravity above $a_0$ at the center, and thus
are the equivalent and do behave like high surface brightness
elliptical galaxies  (e.g., \cite{Carollo95}, \cite{Mehlert00}).

The available velocity dispersion data, however, in no case do probe
radii larger than 2.5$r_0$.  At this radius the constant velocity
dispersion hypothesis predicts a dispersion 60\% higher than Newtonian
dynamics, assuming the two models coincide at $r_0$. This difference
is small enough that taken singularly none of these clusters
represents a compelling case  in favor (or against) a 
Keplerian falloff of the velocity dispersion. Therefore
Newtonian dynamics remaining a viable explanation to describe the data
of individual objects
(e.g., \cite{Moffat08}).  To increase the strength of the signal we
then combine the VDP of all the clusters. Data are normalized plotting
distance in units of $r_0$, and dispersions in units of $\sigma_0$,
the dispersion at $r_0$.  Note that compared to the full cluster size,
$r_0$ corresponds to a very large radius, within which virtually all
the cluster mass is contained.  Thus, no matter what is the exact mass
distribution within the clusters, for $r>r_0$ the dispersion should
follow closely a Keplerian decline (unless external effects alters it).

\begin{figure}
\centering
\includegraphics[height=8cm]{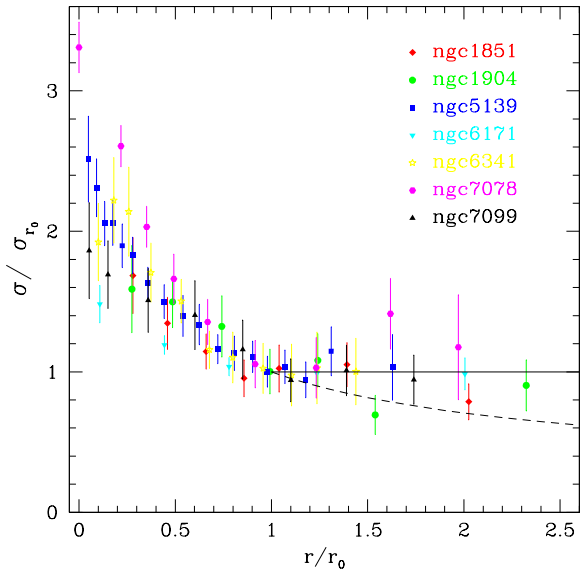}
\caption{\label{dispAllClusters} Normalized velocity dispersion
  profile for all high density clusters studied so far as part of this
  program. Radii are in units of $r_0$, the radius where the
  acceleration is $a_0$, computed assuming M/L=1 for all clusters. The
  dispersion is normalized to the value at $r_0$ to ensure in the 
  region $r>r_0$ profiles from different clusters  are comparable. The
  solid line at $d=1$ represent a constant velocity dispersion, while
  the dashed line gives the Keplerian falloff.  Data for NGC 1851 and
  1904 are from this work, for NGC 5139 from \cite{Scarpa10}, for NGC
  6171 from Scarpa, Marconi \& Gilmozzi 2004A,B, and for NGC 7099 and
  NGC 6343 from \cite{Scarpa07a} }
\end{figure}

The systematic deviation of the VDP from the Newtonian prediction is
now more evident, for all the points but one are above the Keplerian
falloff (Fig. \ref{dispAllClusters}).  To
quantify the difference between the two models we first notice that
they must coincide for $r=r_0$.  Therefore we limit the comparison to
the 16 points at $r>1.2r_0$.  The hypothesis $\sigma/\sigma_0=1$ is
statistically acceptable ($\chi^2=12.0$ for 15 degrees of freedom,
probability 67\%), while a curve falling as $x^{-1/2}$ has
$\chi^2=26.9$ and can be rejected at the 97\% confidence level.

The comparison of a K-S test between the observed data and the 
theoretical models show the constant dispersion hypothesis is 
about 15 times more probable then the Keplerian falloff.

The main conclusion of our work is that all globular clusters exhibits
a flattening of the VDP at large radii.  The explanation for this
phenomenon cannot be the tidal action of the Milky Way, but has to be
something else that applies to all clusters, independently of their
location in the Milky Way halo.

This picture would be in good agreement with the prediction of the
MOND hypothesis, in the case of no strong gravitational external field
(Milgrom 1983). However, according to the MOND original formulation
in none of these clusters we should observe
deviations from Newtonian dynamics, because the external field of the
Milky Way is close or above $a_0$.  Therefore our results appear formally in
disagreement with MOND predictions.

We thanks Y. Momany for providing the photometry of NGC1851,
D. Bettoni for useful comments, and the Italian Space Agency for
economical support by ASI-COFIC contract n. I/016/07/0 "Studi di
Cosmologia e Fisica Fondamentale".

\begin{table}[b]
\label{tabCluster}
\scriptsize
\caption{NGC 1851 and NGC 1904 basic properties from Harris 1996}
\begin{tabular}{lll}
\hline
NGC 1851\\
RA, DEC (2000) &  05:14:06.3 $-40$:02:50  &Coordinates of cluster center\\
L,B            &    244.51  $-35$.04  &Galactic coordinates\\
R$_{sun}$  &   12.1  kpc & Distance from sun\\
R$_{MW}$   &   16.7  kpc & Distance from Milky Way center\\   
V$_{los}$  & 320.5 km/s & Line of sight radial velocity\\
M$_V$ & $-8.33$   &Total V band magnitude\\
Mass/M$\odot$  & 1.8$\times 10^5$ &From luminosity assuming M/L$_V$=1\\ 
r$_h$ &0.52 arcmin or  1.83 pc & Half light radius\\
r$_t$ &11.7 arcmin or 41.2 pc & Tidal radius\\
r$_0$ & 4.1 arcmin or 14.5 pc & Radius where a=$1.2\times 10^{-8}$ cm s$^{-2}$\\
Scale factor & 3.52 &pc/arcmin \\
~\\
NGC 1904\\
RA, DEC (2000) &  05:24:10.6 -24:31:27  &Coordinates of cluster center\\
L,B            &  227.23 -29.53  &Galactic coordinates\\
R$_{sun}$  &   12.9  kpc & Distance from sun\\
R$_{MW}$   &   18.8  kpc & Distance from Milky Way center\\   
V$_{los}$  & 206.0 km/s & Line of sight radial velocity\\
M$_V$ & $-7.86$   &Total V band magnitude\\
Mass/M$\odot$  & 1.2$\times 10^5$ &From luminosity assuming M/L$_V$=1\\ 
r$_h$ &0.80 arcmin or  3.0 pc & Half light radius\\
r$_t$ &8.34 arcmin or 31.3 pc & Tidal radius\\
r$_0$ & 3.1 arcmin or 11.7 pc & Radius where a=$1.2\times 10^{-8}$ cm s$^{-2}$\\
Scale factor & 3.75 &pc/arcmin \\
\hline
\hline
\end{tabular}
\end{table}

\begin{table}[b]
\label{tabVel1851}
\scriptsize
\caption{Radial velocities of stars NGC 1851}
\begin{tabular}{cccccccc}
\# &   ID   &      RA      &     DEC      &          Vel      &   N & V mag\\
   &        &    (2000)    &    (2000)    &         [km/s]    &  \\
\hline
  1 & 14729 &  05:13:23.70 &  -40:05:00.5 &$   320.8 \pm 0.7 $&   5 &  18.8 \\ 
  2 & 13007 &  05:13:29.96 &  -40:11:57.2 &$   313.6 \pm 1.9 $&   3 &  18.5 \\ 
  3 & 14589 &  05:13:29.97 &  -40:05:34.2 &$   321.8 \pm 1.9 $&   5 &  18.7 \\ 
  4 & 15021 &  05:13:30.49 &  -40:04:12.6 &$   312.7 \pm 0.9 $&   5 &  18.5 \\ 
  5 & 15277 &  05:13:32.52 &  -40:03:45.4 &$   318.7 \pm 0.5 $&   5 &  18.1 \\ 
  6 & 14655 &  05:13:35.32 &  -40:05:18.4 &$   315.2 \pm 0.7 $&   5 &  18.4 \\ 
  7 & 14294 &  05:13:35.66 &  -40:06:40.3 &$   319.4 \pm 0.9 $&   5 &  18.2 \\ 
  8 & 14141 &  05:13:37.15 &  -40:07:18.5 &$   328.7 \pm 1.2 $&   4 &  18.9 \\ 
  9 & 14831 &  05:13:37.43 &  -40:04:41.1 &$   322.0 \pm 1.4 $&   4 &  18.6 \\ 
 10 & 14080 &  05:13:38.68 &  -40:07:39.5 &$   318.5 \pm 0.0 $&  10 &  14.3 \\ 
 11 & 16298 &  05:13:40.09 &  -40:00:41.6 &$   320.7 \pm 0.2 $&   5 &  18.2 \\ 
 12 & 16651 &  05:13:40.44 &  -39:59:26.4 &$   318.8 \pm 0.4 $&   5 &  18.7 \\ 
 13 & 13707 &  05:13:41.25 &  -40:09:47.1 &$   314.1 \pm 0.3 $&   5 &  18.2 \\ 
 14 & 15389 &  05:13:43.31 &  -40:03:24.7 &$   312.6 \pm 0.8 $&   4 &  18.8 \\ 
 15 & 19022 &  05:13:46.77 &  -40:09:29.8 &$   320.4 \pm 0.6 $&   5 &  18.4 \\ 
 16 & 19884 &  05:13:46.98 &  -40:07:28.5 &$   312.6 \pm 0.1 $&   2 &  19.0 \\ 
 17 & 24548 &  05:13:47.44 &  -40:04:40.0 &$   321.0 \pm 0.4 $&   5 &  18.5 \\ 
 18 & 19564 &  05:13:47.89 &  -40:08:02.6 &$   320.0 \pm 0.8 $&   5 &  18.9 \\ 
 19 & 31154 &  05:13:48.33 &  -40:03:15.3 &$   327.3 \pm 0.0 $&  10 &  15.5 \\ 
 20 & 19796 &  05:13:48.39 &  -40:07:37.5 &$   323.4 \pm 0.8 $&   5 &  18.8 \\ 
 21 & 39698 &  05:13:48.46 &  -40:01:38.6 &$   318.8 \pm 0.3 $&   5 &  18.3 \\ 
 22 & 21885 &  05:13:48.79 &  -40:05:40.3 &$   316.9 \pm 0.3 $&   5 &  18.7 \\ 
 23 & 46609 &  05:13:48.85 &  -39:59:06.7 &$   322.8 \pm 1.7 $&   4 &  19.0 \\ 
 24 & 30057 &  05:13:50.77 &  -40:03:27.0 &$   322.4 \pm 0.4 $&   5 &  19.0 \\ 
 25 & 29934 &  05:13:50.87 &  -40:03:28.3 &$   312.8 \pm 1.2 $&   4 &  19.0 \\ 
 26 & 19307 &  05:13:51.16 &  -40:08:38.4 &$   318.3 \pm 0.4 $&   5 &  18.8 \\ 
 27 & 26382 &  05:13:51.77 &  -40:04:12.1 &$   310.4 \pm 0.6 $&   5 &  18.7 \\ 
 28 & 21182 &  05:13:51.81 &  -40:06:05.8 &$   316.0 \pm 0.6 $&   5 &  18.9 \\ 
 29 & 27495 &  05:13:51.85 &  -40:03:57.6 &$   320.3 \pm 0.4 $&   5 &  18.8 \\ 
 30 & 18850 &  05:13:52.04 &  -40:10:05.1 &$   315.5 \pm 0.4 $&   5 &  18.4 \\ 
 31 & 20253 &  05:13:52.38 &  -40:06:57.7 &$   323.6 \pm 1.1 $&   5 &  18.7 \\ 
 32 & 27882 &  05:13:52.97 &  -40:03:52.8 &$   319.6 \pm 0.3 $&   5 &  18.2 \\ 
 33 & 43446 &  05:13:53.03 &  -40:00:43.1 &$   319.4 \pm 0.0 $&  10 &  14.1 \\ 
 34 & 31790 &  05:13:53.05 &  -40:03:07.7 &$   318.1 \pm 1.1 $&   5 &  18.8 \\ 
 35 & 46323 &  05:13:53.18 &  -39:59:20.3 &$   313.9 \pm 1.0 $&   5 &  19.0 \\ 
 36 & 19139 &  05:13:53.35 &  -40:09:05.0 &$   322.4 \pm 1.4 $&   5 &  18.8 \\ 
 37 & 22073 &  05:13:53.53 &  -40:05:34.1 &$   319.4 \pm 0.5 $&   5 &  18.7 \\ 
 38 &  3192 &  05:13:53.57 &  -39:55:17.9 &$   320.4 \pm 0.1 $&  10 &  14.1 \\ 
 39 & 48437 &  05:13:53.90 &  -39:55:47.9 &$   317.5 \pm 0.5 $&   5 &  18.4 \\ 
 40 & 21198 &  05:13:53.92 &  -40:06:05.0 &$   312.5 \pm 0.9 $&   4 &  19.0 \\ 
 41 & 33397 &  05:13:54.09 &  -40:02:49.2 &$   327.4 \pm 0.5 $&   5 &  18.3 \\ 
 42 & 33956 &  05:13:54.18 &  -40:02:43.6 &$   328.6 \pm 0.3 $&   5 &  18.1 \\ 
 43 & 21565 &  05:13:54.20 &  -40:05:51.3 &$   317.3 \pm 1.9 $&   3 &  19.0 \\ 
 44 & 28695 &  05:13:54.24 &  -40:03:42.7 &$   320.2 \pm 0.4 $&   5 &  18.8 \\ 
 45 & 37483 &  05:13:54.47 &  -40:02:04.2 &$   313.1 \pm 0.3 $&   5 &  18.3 \\ 
 46 & 39130 &  05:13:55.58 &  -40:01:45.1 &$   326.6 \pm 0.4 $&   5 &  18.8 \\ 
 47 & 23915 &  05:13:55.63 &  -40:04:50.8 &$   325.9 \pm 0.2 $&   5 &  18.2 \\ 
 48 & 45220 &  05:13:55.93 &  -40:00:00.5 &$   322.4 \pm 0.2 $&   5 &  18.2 \\ 
 49 & 46912 &  05:13:56.62 &  -39:58:47.8 &$   320.7 \pm 0.3 $&   5 &  18.0 \\ 
 50 & 31377 &  05:13:56.63 &  -40:03:12.4 &$   319.6 \pm 0.8 $&   5 &  18.9 \\ 
 51 & 43998 &  05:13:56.97 &  -40:00:31.8 &$   319.3 \pm 0.3 $&   5 &  18.4 \\ 
 52 & 41426 &  05:13:57.03 &  -40:01:15.3 &$   317.1 \pm 0.3 $&   5 &  18.4 \\ 
 53 & 36571 &  05:13:57.16 &  -40:02:14.1 &$   327.0 \pm 0.7 $&   5 &  19.0 \\ 
 54 & 44134 &  05:13:57.69 &  -40:00:28.7 &$   322.5 \pm 0.5 $&   5 &  18.8 \\ 
 55 & 38261 &  05:13:57.71 &  -40:01:55.4 &$   324.5 \pm 0.5 $&   5 &  18.7 \\ 
 56 & 30298 &  05:13:57.88 &  -40:03:24.4 &$   323.5 \pm 0.3 $&   5 &  18.5 \\ 
 57 & 27236 &  05:13:58.13 &  -40:04:00.6 &$   312.8 \pm 0.5 $&   5 &  18.7 \\ 
 58 & 19324 &  05:13:58.24 &  -40:08:34.7 &$   327.6 \pm 0.7 $&   5 &  18.9 \\ 
 59 & 22905 &  05:13:58.26 &  -40:05:11.7 &$   313.7 \pm 0.5 $&   5 &  19.0 \\ 
 60 & 19443 &  05:13:58.78 &  -40:08:19.0 &$   312.6 \pm 0.3 $&   5 &  18.5 \\ 
 61 & 32453 &  05:13:59.00 &  -40:02:59.9 &$   314.9 \pm 0.2 $&   5 &  18.2 \\ 
 62 & 34013 &  05:13:59.08 &  -40:02:42.8 &$   323.4 \pm 0.3 $&   5 &  18.4 \\ 
 63 & 22212 &  05:13:59.12 &  -40:05:30.2 &$   314.6 \pm 0.3 $&   5 &  18.5 \\ 
 64 & 35699 &  05:13:59.29 &  -40:02:24.1 &$   311.6 \pm 0.4 $&   5 &  18.5 \\ 
 65 & 24570 &  05:13:59.47 &  -40:04:39.3 &$   319.3 \pm 0.4 $&   5 &  18.9 \\ 
 66 & 22741 &  05:14:00.05 &  -40:05:15.8 &$   327.6 \pm 1.1 $&   5 &  18.8 \\ 
 67 & 41045 &  05:14:00.19 &  -40:01:20.8 &$   313.4 \pm 0.1 $&   5 &  18.3 \\ 
 68 & 47835 &  05:14:00.69 &  -39:57:20.0 &$   323.0 \pm 0.8 $&   5 &  18.9 \\ 
 69 & 34266 &  05:14:00.72 &  -40:02:40.1 &$   314.9 \pm 0.3 $&   5 &  18.1 \\ 
 70 & 21112 &  05:14:00.96 &  -40:06:08.1 &$   323.6 \pm 0.4 $&   5 &  19.0 \\ 
 71 & 40916 &  05:14:01.03 &  -40:01:22.4 &$   325.0 \pm 0.4 $&   5 &  18.2 \\ 
 72 & 33116 &  05:14:01.05 &  -40:02:52.0 &$   324.6 \pm 0.6 $&   5 &  19.0 \\ 
 73 & 31177 &  05:14:01.24 &  -40:03:14.7 &$   313.8 \pm 0.6 $&   5 &  18.6 \\ 
 74 & 28311 &  05:14:01.76 &  -40:03:47.2 &$   327.5 \pm 0.2 $&   5 &  18.0 \\ 
 75 & 20826 &  05:14:02.73 &  -40:06:21.2 &$   317.2 \pm 0.4 $&   3 &  19.0 \\ 
 76 & 47497 &  05:14:03.12 &  -39:57:57.2 &$   319.3 \pm 0.4 $&   5 &  18.2 \\ 
 77 & 20350 &  05:14:03.17 &  -40:06:50.6 &$   318.7 \pm 1.3 $&   5 &  18.9 \\ 
 78 & 37003 &  05:14:03.21 &  -40:02:09.3 &$   317.3 \pm 0.2 $&   5 &  18.3 \\ 
 79 & 23472 &  05:14:03.37 &  -40:04:59.0 &$   313.4 \pm 0.4 $&   5 &  18.1 \\ 
 80 & 21134 &  05:14:03.86 &  -40:06:07.3 &$   320.7 \pm 0.3 $&   5 &  18.6 \\ 
\hline
\hline
\end{tabular}
\end{table}
\addtocounter{table}{-1}

\begin{table}[b]
\scriptsize
\caption{Radial velocities for NGC 1851 -- continued}
\begin{tabular}{cccccccc}
\# &   ID   &      RA      &     DEC      &          Vel      &   N & V mag\\
   &        &    (2000)    &    (2000)    &         [km/s]    &  \\
\hline
 81 & 31131 &  05:14:03.86 &  -40:03:15.0 &$   324.8 \pm 0.5 $&   5 &  18.7 \\ 
 82 & 27070 &  05:14:03.93 &  -40:04:02.7 &$   311.8 \pm 0.3 $&   5 &  18.2 \\ 
 83 & 42138 &  05:14:04.00 &  -40:01:04.6 &$   315.3 \pm 0.3 $&   5 &  18.5 \\ 
 84 & 19554 &  05:14:04.05 &  -40:08:03.5 &$   316.2 \pm 0.8 $&   5 &  19.0 \\ 
 85 & 22153 &  05:14:04.07 &  -40:05:31.8 &$   320.0 \pm 0.6 $&   5 &  18.5 \\ 
 86 & 45651 &  05:14:04.07 &  -39:59:46.4 &$   312.3 \pm 0.1 $&   5 &  18.6 \\ 
 87 & 39581 &  05:14:04.30 &  -40:01:39.7 &$   318.3 \pm 0.2 $&   5 &  18.0 \\ 
 88 & 42097 &  05:14:04.32 &  -40:01:05.1 &$   325.8 \pm 0.1 $&   5 &  18.1 \\ 
 89 & 39271 &  05:14:04.65 &  -40:01:43.2 &$   308.3 \pm 0.4 $&   5 &  18.0 \\ 
 90 & 46822 &  05:14:04.69 &  -39:58:53.4 &$   319.2 \pm 0.3 $&   5 &  18.5 \\ 
 91 & 27204 &  05:14:04.87 &  -40:04:01.0 &$   318.1 \pm 1.0 $&   5 &  18.9 \\ 
 92 & 22677 &  05:14:04.91 &  -40:05:17.2 &$   324.7 \pm 0.7 $&   5 &  18.8 \\ 
 93 & 48120 &  05:14:05.03 &  -39:56:36.4 &$    43.5 \pm 0.3 $&   5 &  18.4 \\ 
 94 & 42745 &  05:14:05.43 &  -40:00:54.7 &$   314.6 \pm 0.4 $&   5 &  18.4 \\ 
 95 & 40502 &  05:14:05.52 &  -40:01:27.7 &$   311.8 \pm 0.5 $&   5 &  18.5 \\ 
 96 & 28639 &  05:14:05.63 &  -40:03:43.2 &$   308.3 \pm 0.4 $&   5 &  18.4 \\ 
 97 & 46458 &  05:14:05.97 &  -39:59:13.5 &$   318.4 \pm 0.3 $&   5 &  18.0 \\ 
 98 & 21828 &  05:14:05.98 &  -40:05:41.9 &$   312.4 \pm 0.4 $&   5 &  18.8 \\ 
 99 & 21165 &  05:14:06.00 &  -40:06:06.3 &$   328.7 \pm 0.3 $&   5 &  18.9 \\ 
100 & 19309 &  05:14:06.94 &  -40:08:37.8 &$   317.8 \pm 0.8 $&   5 &  19.0 \\ 
101 & 45709 &  05:14:07.16 &  -39:59:44.3 &$   330.2 \pm 0.2 $&   5 &  18.8 \\ 
102 & 26167 &  05:14:07.35 &  -40:04:14.7 &$   319.4 \pm 0.2 $&   5 &  18.5 \\ 
103 & 46644 &  05:14:07.44 &  -39:59:04.4 &$   316.8 \pm 1.4 $&   5 &  19.0 \\ 
104 & 45848 &  05:14:07.50 &  -39:59:39.3 &$   318.1 \pm 0.5 $&   5 &  18.5 \\ 
105 & 28825 &  05:14:07.59 &  -40:03:40.6 &$   332.1 \pm 0.4 $&   5 &  18.4 \\ 
106 & 38957 &  05:14:07.79 &  -40:01:47.2 &$   329.7 \pm 0.2 $&   5 &  18.3 \\ 
107 & 19805 &  05:14:08.17 &  -40:07:36.0 &$   322.5 \pm 0.4 $&   5 &  18.1 \\ 
108 & 22255 &  05:14:08.57 &  -40:05:28.5 &$   320.5 \pm 0.5 $&   5 &  18.9 \\ 
109 & 19200 &  05:14:08.81 &  -40:08:54.8 &$   314.8 \pm 1.2 $&   4 &  18.9 \\ 
110 & 20851 &  05:14:08.92 &  -40:06:19.5 &$   318.2 \pm 0.4 $&   5 &  18.7 \\ 
111 & 46908 &  05:14:08.94 &  -39:58:47.9 &$   316.5 \pm 0.2 $&   5 &  18.3 \\ 
112 & 46096 &  05:14:08.94 &  -39:59:30.1 &$   314.9 \pm 0.5 $&   5 &  19.0 \\ 
113 & 44889 &  05:14:08.96 &  -40:00:09.4 &$   320.5 \pm 0.4 $&   5 &  18.1 \\ 
114 & 46433 &  05:14:09.16 &  -39:59:14.5 &$   315.2 \pm 1.0 $&   4 &  19.0 \\ 
115 & 18656 &  05:14:09.71 &  -40:10:44.6 &$   315.6 \pm 1.0 $&   5 &  18.4 \\ 
116 & 40705 &  05:14:09.84 &  -40:01:24.8 &$   330.4 \pm 0.1 $&   5 &  18.3 \\ 
117 & 39847 &  05:14:09.86 &  -40:01:36.2 &$   313.9 \pm 0.0 $&   5 &  13.9 \\ 
118 & 39006 &  05:14:10.11 &  -40:01:46.4 &$   324.1 \pm 0.3 $&   5 &  18.2 \\ 
119 & 29582 &  05:14:10.23 &  -40:03:31.7 &$   326.0 \pm 1.1 $&   5 &  18.2 \\ 
120 & 41432 &  05:14:10.73 &  -40:01:14.9 &$   321.4 \pm 0.5 $&   5 &  18.8 \\ 
121 & 48457 &  05:14:11.04 &  -39:55:43.9 &$   317.3 \pm 1.1 $&   5 &  19.0 \\ 
122 & 41502 &  05:14:11.23 &  -40:01:13.7 &$   313.6 \pm 0.4 $&   5 &  18.8 \\ 
123 & 20559 &  05:14:11.34 &  -40:06:37.4 &$   315.7 \pm 0.7 $&   3 &  19.0 \\ 
124 & 38347 &  05:14:11.90 &  -40:01:53.9 &$   328.8 \pm 0.5 $&   5 &  18.9 \\ 
125 & 43953 &  05:14:12.04 &  -40:00:32.3 &$   319.7 \pm 0.6 $&   5 &  19.0 \\ 
126 & 38826 &  05:14:12.29 &  -40:01:48.7 &$   327.3 \pm 0.8 $&   5 &  18.7 \\ 
127 & 47224 &  05:14:12.29 &  -39:58:24.7 &$   321.4 \pm 0.1 $&   5 &  18.8 \\ 
128 & 23057 &  05:14:12.35 &  -40:05:07.5 &$   319.5 \pm 0.4 $&   5 &  18.5 \\ 
129 & 25840 &  05:14:12.46 &  -40:04:19.1 &$   316.9 \pm 0.4 $&   5 &  18.9 \\ 
130 & 45531 &  05:14:12.94 &  -39:59:50.3 &$   328.1 \pm 1.0 $&   5 &  19.0 \\ 
131 & 22278 &  05:14:13.00 &  -40:05:27.7 &$   319.8 \pm 0.6 $&   5 &  18.7 \\ 
132 & 28104 &  05:14:13.10 &  -40:03:49.6 &$   309.0 \pm 0.2 $&   5 &  19.0 \\ 
133 & 40619 &  05:14:13.54 &  -40:01:25.7 &$   314.4 \pm 0.1 $&   5 &  18.3 \\ 
134 & 21810 &  05:14:13.86 &  -40:05:42.1 &$   318.0 \pm 0.3 $&   5 &  18.7 \\ 
135 & 46033 &  05:14:14.28 &  -39:59:32.3 &$   314.9 \pm 0.8 $&   4 &  19.0 \\ 
136 & 41419 &  05:14:14.72 &  -40:01:14.9 &$   326.9 \pm 0.5 $&   5 &  19.0 \\ 
137 & 18150 &  05:14:14.78 &  -40:11:38.9 &$   320.4 \pm 0.6 $&   5 &  18.0 \\ 
138 & 42970 &  05:14:15.08 &  -40:00:50.7 &$   317.4 \pm 0.1 $&   5 &  18.2 \\ 
139 & 26485 &  05:14:15.08 &  -40:04:09.9 &$   320.2 \pm 0.6 $&   5 &  19.0 \\ 
140 & 39875 &  05:14:15.17 &  -40:01:35.7 &$   323.3 \pm 0.5 $&   4 &  18.5 \\ 
141 & 25964 &  05:14:15.48 &  -40:04:17.2 &$   322.5 \pm 0.6 $&   5 &  18.9 \\ 
142 & 27421 &  05:14:15.57 &  -40:03:57.8 &$   326.2 \pm 0.4 $&   5 &  18.6 \\ 
143 & 21484 &  05:14:15.80 &  -40:05:53.4 &$   324.1 \pm 1.1 $&   4 &  19.0 \\ 
144 & 46479 &  05:14:15.84 &  -39:59:12.1 &$   326.4 \pm 0.3 $&   5 &  18.5 \\ 
145 & 17879 &  05:14:16.25 &  -40:11:57.1 &$   318.1 \pm 0.8 $&   4 &  18.9 \\ 
146 & 42908 &  05:14:16.29 &  -40:00:51.8 &$   318.9 \pm 0.2 $&   5 &  18.2 \\ 
147 & 25605 &  05:14:16.59 &  -40:04:22.3 &$   323.9 \pm 0.6 $&   5 &  19.0 \\ 
148 & 43709 &  05:14:16.75 &  -40:00:37.0 &$   324.6 \pm 0.9 $&   5 &  19.0 \\ 
149 & 40153 &  05:14:16.76 &  -40:01:32.0 &$   324.6 \pm 0.4 $&   5 &  19.0 \\ 
150 & 33925 &  05:14:16.77 &  -40:02:43.2 &$   318.8 \pm 0.3 $&   5 &  18.1 \\ 
151 & 44016 &  05:14:16.92 &  -40:00:30.7 &$   323.2 \pm 0.9 $&   4 &  19.0 \\ 
152 & 26933 &  05:14:16.96 &  -40:04:03.9 &$   315.9 \pm 0.4 $&   5 &  18.8 \\ 
153 & 42256 &  05:14:17.06 &  -40:01:02.2 &$   329.0 \pm 1.2 $&   5 &  18.9 \\ 
154 & 46284 &  05:14:17.90 &  -39:59:21.3 &$   322.0 \pm 0.4 $&   5 &  18.4 \\ 
155 & 31495 &  05:14:17.91 &  -40:03:10.4 &$   320.4 \pm 0.3 $&   4 &  19.0 \\ 
156 & 45371 &  05:14:17.99 &  -39:59:55.4 &$   315.3 \pm 0.3 $&   5 &  18.0 \\ 
157 & 37497 &  05:14:18.08 &  -40:02:03.2 &$   316.1 \pm 0.3 $&   5 &  18.3 \\ 
158 & 26282 &  05:14:18.74 &  -40:04:12.7 &$   321.1 \pm 0.4 $&   5 &  18.7 \\ 
159 & 29424 &  05:14:19.08 &  -40:03:33.2 &$   321.7 \pm 0.1 $&   5 &  18.2 \\ 
160 & 42103 &  05:14:19.55 &  -40:01:04.5 &$   325.4 \pm 0.2 $&   5 &  18.9 \\ 
\hline
\hline
\end{tabular}
\end{table}
\addtocounter{table}{-1}

\begin{table}[b]
\scriptsize
\caption{Radial velocities of stars in NGC 1851 -- continued}
\begin{tabular}{cccccccc}
\# &   ID   &      RA      &     DEC      &          Vel      &   N & V mag\\
   &        &    (2000)    &    (2000)    &         [km/s]    &  \\
\hline
161 & 46497 &  05:14:20.25 &  -39:59:11.0 &$   320.0 \pm 0.3 $&   5 &  18.4 \\ 
162 & 19441 &  05:14:20.71 &  -40:08:18.5 &$   321.7 \pm 0.5 $&   5 &  18.6 \\ 
163 & 35357 &  05:14:20.72 &  -40:02:27.4 &$   324.4 \pm 0.2 $&   5 &  18.1 \\ 
164 & 40709 &  05:14:20.83 &  -40:01:24.3 &$   320.7 \pm 0.6 $&   5 &  18.9 \\ 
165 & 20506 &  05:14:20.91 &  -40:06:39.8 &$   325.5 \pm 0.4 $&   5 &  18.6 \\ 
166 & 30021 &  05:14:21.10 &  -40:03:26.5 &$   319.3 \pm 0.1 $&   4 &  18.1 \\ 
167 & 45097 &  05:14:21.40 &  -40:00:02.9 &$   320.7 \pm 0.4 $&   5 &  18.2 \\ 
168 & 19664 &  05:14:21.70 &  -40:07:49.9 &$   320.5 \pm 0.6 $&   5 &  18.7 \\ 
169 & 34410 &  05:14:22.03 &  -40:02:37.7 &$   313.2 \pm 0.2 $&   5 &  18.9 \\ 
170 & 44618 &  05:14:22.11 &  -40:00:16.4 &$   321.8 \pm 0.1 $&   5 &  18.2 \\ 
171 & 24751 &  05:14:22.18 &  -40:04:35.3 &$   323.0 \pm 0.2 $&   5 &  18.1 \\ 
172 & 27965 &  05:14:22.43 &  -40:03:51.0 &$   322.3 \pm 0.9 $&   5 &  18.8 \\ 
173 & 46681 &  05:14:22.56 &  -39:59:01.7 &$   316.0 \pm 0.3 $&   5 &  18.7 \\ 
174 & 28310 &  05:14:22.91 &  -40:03:46.4 &$   325.6 \pm 0.3 $&   5 &  18.6 \\ 
175 & 22954 &  05:14:23.07 &  -40:05:09.8 &$   315.8 \pm 0.5 $&   5 &  18.6 \\ 
176 & 40476 &  05:14:23.85 &  -40:01:27.4 &$   326.3 \pm 0.7 $&   5 &  18.9 \\ 
177 &  3799 &  05:14:24.22 &  -39:52:30.3 &$   321.8 \pm 0.8 $&   5 &  18.9 \\ 
178 &  3717 &  05:14:24.44 &  -39:53:12.2 &$   320.7 \pm 0.7 $&   5 &  18.5 \\ 
179 & 24263 &  05:14:24.82 &  -40:04:43.8 &$   316.5 \pm 0.4 $&   5 &  18.4 \\ 
180 & 20919 &  05:14:25.10 &  -40:06:15.2 &$   321.1 \pm 0.1 $&   5 &  18.8 \\ 
181 & 24840 &  05:14:25.16 &  -40:04:33.5 &$   320.0 \pm 0.4 $&   5 &  18.8 \\ 
182 & 41815 &  05:14:25.32 &  -40:01:08.6 &$   317.5 \pm 0.9 $&   5 &  18.9 \\ 
183 & 32903 &  05:14:25.97 &  -40:02:53.8 &$   319.7 \pm 0.1 $&  10 &  13.8 \\ 
184 & 22338 &  05:14:26.21 &  -40:05:25.6 &$   322.8 \pm 0.4 $&   5 &  18.3 \\ 
185 & 45391 &  05:14:26.72 &  -39:59:54.5 &$   310.0 \pm 0.5 $&   5 &  18.5 \\ 
186 &  3498 &  05:14:27.29 &  -39:54:24.5 &$   322.6 \pm 0.4 $&   5 &  18.2 \\ 
187 & 34212 &  05:14:27.64 &  -40:02:39.6 &$   328.5 \pm 0.5 $&   5 &  18.2 \\ 
188 & 54310 &  05:14:32.11 &  -39:57:12.7 &$    90.2 \pm 0.2 $&   5 &  18.2 \\ 
189 & 53852 &  05:14:32.13 &  -39:58:46.3 &$   326.1 \pm 0.5 $&   5 &  18.8 \\ 
190 & 49277 &  05:14:33.22 &  -40:11:13.8 &$   319.4 \pm 0.5 $&   5 &  18.6 \\ 
191 & 52750 &  05:14:34.75 &  -40:01:31.5 &$   322.9 \pm 0.6 $&   4 &  18.9 \\ 
192 & 52972 &  05:14:34.88 &  -40:00:58.6 &$   323.0 \pm 0.4 $&   5 &  18.8 \\ 
193 & 52569 &  05:14:36.07 &  -40:01:54.7 &$   326.6 \pm 0.5 $&   5 &  18.9 \\ 
194 & 51626 &  05:14:36.81 &  -40:03:59.1 &$   322.2 \pm 1.5 $&   5 &  18.8 \\ 
195 & 50507 &  05:14:37.48 &  -40:07:04.2 &$   316.1 \pm 0.3 $&   5 &  18.3 \\ 
196 & 50690 &  05:14:45.87 &  -40:06:25.4 &$    56.7 \pm 0.2 $&   3 &  19.0 \\ 
197 & 52758 &  05:14:48.74 &  -40:01:29.5 &$   318.1 \pm 0.6 $&   3 &  19.0 \\ 
198 & 51770 &  05:14:56.62 &  -40:03:37.7 &$   318.7 \pm 0.5 $&   5 &  18.4 \\ 
199 & 52120 &  05:15:04.47 &  -40:02:51.8 &$    80.8 \pm 0.8 $&   4 &  18.6 \\ 
\hline
\hline
\end{tabular}
\end{table}

\begin{table}[b]
\label{tabVel1904}
\scriptsize
\caption{Radial velocities of stars in NGC 1904}
\begin{tabular}{cccccccc}
\# & ID &   RA   &  DEC   & Vel   &   N     & V mag\\
   &    & (2000) & (2000) & [km/s]    & & \\
\hline
  1 &  1179 &  05:23:29.73 &  -24:30:26.4 &$    35.3 \pm 0.1 $&  10 &  17.5 \\ 
  2 &  1443 &  05:23:30.75 &  -24:33:32.3 &$   200.0 \pm 0.8 $&   5 &  17.9 \\ 
  3 &  2177 &  05:23:32.06 &  -24:27:12.5 &$    33.5 \pm 1.2 $&   4 &  18.6 \\ 
  4 &  2392 &  05:23:41.47 &  -24:30:35.7 &$   206.8 \pm 0.7 $&   5 &  18.8 \\ 
  5 &  2678 &  05:23:42.48 &  -24:29:01.3 &$   203.1 \pm 1.1 $&   4 &  19.0 \\ 
  6 &  1370 &  05:23:43.53 &  -24:32:43.7 &$    44.6 \pm 2.8 $&   5 &  17.7 \\ 
  7 &  2747 &  05:23:44.08 &  -24:29:32.9 &$   204.7 \pm 1.2 $&   3 &  19.0 \\ 
  8 &  2609 &  05:23:44.11 &  -24:24:29.2 &$   100.9 \pm 0.7 $&   4 &  18.3 \\ 
  9 &  1881 &  05:23:44.62 &  -24:36:41.3 &$   203.3 \pm 0.5 $&   5 &  18.4 \\ 
 10 &  1930 &  05:23:44.72 &  -24:30:35.0 &$   206.1 \pm 0.3 $&   5 &  18.4 \\ 
 11 &  2215 &  05:23:45.15 &  -24:34:52.6 &$    44.6 \pm 0.1 $&   5 &  18.5 \\ 
 12 &  2708 &  05:23:45.66 &  -24:30:07.1 &$   203.8 \pm 0.5 $&   3 &  19.1 \\ 
 13 &  2057 &  05:23:48.72 &  -24:31:30.4 &$   204.5 \pm 0.4 $&   5 &  18.6 \\ 
 14 &  1569 &  05:23:48.78 &  -24:32:19.4 &$   204.0 \pm 0.3 $&   5 &  18.1 \\ 
 15 &  1727 &  05:23:51.26 &  -24:31:09.9 &$   206.5 \pm 0.2 $&   5 &  18.3 \\ 
 16 &  1319 &  05:23:51.58 &  -24:31:50.7 &$   206.5 \pm 0.1 $&   5 &  17.7 \\ 
 17 &  1894 &  05:23:52.45 &  -24:31:36.2 &$   209.3 \pm 0.2 $&   5 &  18.4 \\ 
 18 &  2158 &  05:23:53.39 &  -24:31:29.7 &$   203.9 \pm 0.5 $&   3 &  18.6 \\ 
 19 &  1954 &  05:23:56.37 &  -24:30:19.0 &$   208.0 \pm 0.3 $&   5 &  18.4 \\ 
 20 &  1815 &  05:23:56.56 &  -24:29:52.9 &$   208.6 \pm 0.3 $&   5 &  18.3 \\ 
 21 &  2837 &  05:23:56.58 &  -24:30:37.3 &$   207.9 \pm 1.0 $&   5 &  19.1 \\ 
 22 &  1484 &  05:23:56.70 &  -24:33:01.1 &$   209.9 \pm 0.2 $&   5 &  17.9 \\ 
 23 &  1570 &  05:23:57.08 &  -24:29:26.3 &$   207.9 \pm 0.2 $&   5 &  18.0 \\ 
 24 &  1583 &  05:23:57.64 &  -24:29:58.1 &$   208.2 \pm 0.3 $&   5 &  18.0 \\ 
 25 &  1047 &  05:23:58.27 &  -24:33:40.8 &$   200.7 \pm 0.1 $&   5 &  17.3 \\ 
 26 &  1857 &  05:23:58.38 &  -24:31:22.9 &$   206.8 \pm 0.3 $&   5 &  18.4 \\ 
 27 &  1818 &  05:23:58.99 &  -24:29:14.6 &$   207.3 \pm 0.2 $&   5 &  18.3 \\ 
 28 &  1302 &  05:23:59.33 &  -24:29:44.3 &$   208.5 \pm 0.1 $&   5 &  17.6 \\ 
 29 &  2108 &  05:23:59.58 &  -24:35:05.6 &$   206.7 \pm 0.7 $&   5 &  18.5 \\ 
 30 &  1275 &  05:23:59.84 &  -24:31:33.5 &$   206.5 \pm 0.1 $&   5 &  17.6 \\ 
 31 &  1407 &  05:24:00.20 &  -24:30:29.9 &$   206.3 \pm 0.2 $&   5 &  17.8 \\ 
 32 &  1515 &  05:24:00.29 &  -24:30:36.5 &$   206.5 \pm 0.4 $&   5 &  18.0 \\ 
 33 &  2105 &  05:24:00.40 &  -24:34:38.1 &$   204.1 \pm 0.4 $&   5 &  18.5 \\ 
 34 &  2568 &  05:24:00.95 &  -24:31:26.1 &$   206.2 \pm 1.7 $&   5 &  19.0 \\ 
 35 &  2580 &  05:24:01.06 &  -24:29:31.6 &$   206.7 \pm 1.8 $&   5 &  19.0 \\ 
 36 &  2561 &  05:24:01.09 &  -24:29:58.7 &$   207.8 \pm 0.5 $&   5 &  18.8 \\ 
 37 &  1286 &  05:24:01.19 &  -24:29:01.8 &$   210.9 \pm 0.3 $&   5 &  17.7 \\ 
 38 &  2820 &  05:24:01.83 &  -24:30:03.8 &$   204.2 \pm 1.0 $&   5 &  19.0 \\ 
 39 &  2616 &  05:24:02.49 &  -24:32:59.8 &$   199.6 \pm 0.8 $&   5 &  18.9 \\ 
 40 &  1467 &  05:24:02.58 &  -24:31:30.6 &$   202.9 \pm 0.2 $&   5 &  17.9 \\ 
 41 &  1175 &  05:24:03.30 &  -24:32:28.9 &$   202.0 \pm 0.1 $&   5 &  17.5 \\ 
 42 &  2748 &  05:24:03.53 &  -24:27:23.9 &$   208.8 \pm 0.6 $&   4 &  19.1 \\ 
 43 &  2042 &  05:24:04.19 &  -24:30:45.6 &$   209.4 \pm 0.3 $&   5 &  18.5 \\ 
 44 &  2488 &  05:24:04.35 &  -24:30:29.0 &$   210.4 \pm 0.5 $&   5 &  18.8 \\ 
 45 &   949 &  05:24:04.44 &  -24:32:23.0 &$   206.3 \pm 0.1 $&  10 &  17.5 \\ 
 46 &  1362 &  05:24:04.46 &  -24:29:01.8 &$   203.7 \pm 0.3 $&   5 &  17.8 \\ 
 47 &  1028 &  05:24:04.66 &  -24:31:24.1 &$   200.9 \pm 0.2 $&   5 &  17.2 \\ 
 48 &  2614 &  05:24:05.08 &  -24:24:45.6 &$   207.8 \pm 0.1 $&   3 &  19.0 \\ 
 49 &  1525 &  05:24:05.43 &  -24:29:55.2 &$   213.6 \pm 0.2 $&   5 &  18.0 \\ 
 50 &  1306 &  05:24:05.52 &  -24:30:06.7 &$   202.2 \pm 0.1 $&   5 &  17.7 \\ 
 51 &  2686 &  05:24:05.54 &  -24:31:53.1 &$   209.1 \pm 0.3 $&   5 &  18.9 \\ 
 52 &  2462 &  05:24:05.61 &  -24:26:58.5 &$   207.3 \pm 0.6 $&   5 &  18.8 \\ 
 53 &  1916 &  05:24:06.00 &  -24:32:25.6 &$   206.9 \pm 0.4 $&   5 &  18.6 \\ 
 54 &  1311 &  05:24:06.24 &  -24:35:01.5 &$   204.2 \pm 0.2 $&   5 &  17.7 \\ 
 55 &  2476 &  05:24:06.42 &  -24:29:53.5 &$   203.2 \pm 0.6 $&   5 &  19.0 \\ 
 56 &  2554 &  05:24:06.44 &  -24:30:33.0 &$   208.8 \pm 0.5 $&   5 &  19.1 \\ 
 57 &  1322 &  05:24:06.47 &  -24:31:48.0 &$   208.8 \pm 0.3 $&   5 &  17.8 \\ 
 58 &  1099 &  05:24:06.65 &  -24:33:11.2 &$   208.2 \pm 0.3 $&   5 &  17.3 \\ 
 59 &  1080 &  05:24:07.01 &  -24:31:52.0 &$   206.9 \pm 0.1 $&   5 &  17.2 \\ 
 60 &  1474 &  05:24:07.34 &  -24:37:58.0 &$   205.4 \pm 0.4 $&   5 &  17.9 \\ 
 61 &  2294 &  05:24:07.39 &  -24:30:51.3 &$   209.9 \pm 0.4 $&   5 &  18.9 \\ 
 62 &  2688 &  05:24:07.63 &  -24:27:18.0 &$   205.8 \pm 1.0 $&   5 &  19.0 \\ 
 63 &  1408 &  05:24:07.73 &  -24:25:13.3 &$    19.0 \pm 0.1 $&   5 &  17.4 \\ 
 64 &  1222 &  05:24:07.84 &  -24:28:07.7 &$   205.4 \pm 0.2 $&   5 &  17.6 \\ 
 65 &  1388 &  05:24:07.84 &  -24:32:48.7 &$   202.4 \pm 0.2 $&   5 &  17.8 \\ 
 66 &  2302 &  05:24:07.92 &  -24:30:42.5 &$   208.7 \pm 0.4 $&   5 &  18.7 \\ 
 67 &  1258 &  05:24:07.93 &  -24:32:54.8 &$   197.1 \pm 0.2 $&   5 &  17.6 \\ 
 68 &   999 &  05:24:08.04 &  -24:31:21.1 &$   212.6 \pm 0.2 $&   5 &  17.1 \\ 
 69 &  1479 &  05:24:08.37 &  -24:30:17.8 &$   211.3 \pm 0.3 $&   5 &  17.9 \\ 
 70 &  1464 &  05:24:08.44 &  -24:35:37.2 &$   208.9 \pm 0.2 $&   5 &  17.9 \\ 
 71 &  1241 &  05:24:08.54 &  -24:31:58.6 &$   209.7 \pm 0.1 $&   5 &  17.7 \\ 
 72 &  1375 &  05:24:08.64 &  -24:32:20.7 &$   202.1 \pm 0.2 $&   5 &  17.8 \\ 
 73 &  1536 &  05:24:08.67 &  -24:26:25.8 &$   207.4 \pm 0.2 $&   5 &  18.0 \\ 
 74 &  1270 &  05:24:09.12 &  -24:32:33.8 &$   204.3 \pm 0.2 $&   5 &  17.7 \\ 
 75 &  1048 &  05:24:09.16 &  -24:28:46.9 &$   205.7 \pm 0.2 $&   5 &  17.2 \\ 
 76 &  2086 &  05:24:09.16 &  -24:35:02.1 &$   207.7 \pm 0.2 $&   5 &  18.5 \\ 
 77 &  2296 &  05:24:09.32 &  -24:30:38.7 &$   200.5 \pm 1.1 $&   5 &  19.1 \\ 
 78 &  1663 &  05:24:09.38 &  -24:26:01.4 &$   209.3 \pm 0.4 $&   5 &  18.2 \\ 
 79 &  1435 &  05:24:09.42 &  -24:23:11.6 &$   204.4 \pm 0.4 $&   5 &  17.9 \\ 
 80 &  1628 &  05:24:09.44 &  -24:36:34.5 &$   205.6 \pm 0.2 $&   5 &  18.1 \\ 
\hline
\hline
\end{tabular}
\end{table}
\addtocounter{table}{-1}

\begin{table}[b]
\scriptsize
\caption{Radial velocities of stars in NGC 1904 -- continued}
\begin{tabular}{cccccccc}
\# & ID &   RA   &  DEC   & Vel   &   N     & V mag\\
   &    & (2000) & (2000) & [km/s]    & & \\
\hline
 81 &  1239 &  05:24:09.50 &  -24:32:21.3 &$   203.1 \pm 0.1 $&   5 &  17.6 \\ 
 82 &  1759 &  05:24:09.52 &  -24:27:56.1 &$   204.6 \pm 0.5 $&   5 &  18.2 \\ 
 83 &  1503 &  05:24:09.53 &  -24:32:56.7 &$   205.9 \pm 0.2 $&   5 &  17.9 \\ 
 84 &  1372 &  05:24:10.22 &  -24:29:07.5 &$   211.8 \pm 0.2 $&   5 &  17.8 \\ 
 85 &  1238 &  05:24:10.65 &  -24:33:48.6 &$   206.9 \pm 0.2 $&   5 &  17.5 \\ 
 86 &  2569 &  05:24:10.72 &  -24:25:58.8 &$   206.6 \pm 1.2 $&   4 &  18.9 \\ 
 87 &  1505 &  05:24:10.77 &  -24:34:13.9 &$   203.0 \pm 0.4 $&   5 &  17.9 \\ 
 88 &  1482 &  05:24:11.40 &  -24:33:24.7 &$   205.1 \pm 0.3 $&   5 &  17.9 \\ 
 89 &  1609 &  05:24:11.42 &  -24:29:55.2 &$   207.8 \pm 0.3 $&   5 &  18.1 \\ 
 90 &  2911 &  05:24:11.48 &  -24:30:11.0 &$   205.2 \pm 0.5 $&   5 &  19.1 \\ 
 91 &  1007 &  05:24:11.52 &  -24:32:22.2 &$   204.4 \pm 0.1 $&   5 &  17.1 \\ 
 92 &  1779 &  05:24:12.02 &  -24:33:56.3 &$   205.5 \pm 0.2 $&   5 &  18.2 \\ 
 93 &  1353 &  05:24:12.03 &  -24:34:49.4 &$   205.0 \pm 0.3 $&   5 &  17.7 \\ 
 94 &  2485 &  05:24:12.05 &  -24:29:28.6 &$   206.9 \pm 1.2 $&   4 &  19.1 \\ 
 95 &  1428 &  05:24:12.21 &  -24:30:21.5 &$   209.5 \pm 0.1 $&   5 &  17.9 \\ 
 96 &  1194 &  05:24:12.33 &  -24:31:21.8 &$   205.1 \pm 0.3 $&   5 &  17.5 \\ 
 97 &   970 &  05:24:12.34 &  -24:30:33.0 &$   201.5 \pm 0.4 $&   5 &  17.1 \\ 
 98 &  1298 &  05:24:12.45 &  -24:29:39.0 &$   205.7 \pm 0.2 $&   5 &  17.7 \\ 
 99 &  2027 &  05:24:12.54 &  -24:28:44.4 &$   210.0 \pm 0.4 $&   5 &  18.5 \\ 
100 &  1303 &  05:24:12.57 &  -24:32:24.3 &$   199.7 \pm 0.2 $&   5 &  17.8 \\ 
101 &  1109 &  05:24:12.88 &  -24:31:06.6 &$   201.2 \pm 0.1 $&   5 &  17.3 \\ 
102 &  1411 &  05:24:13.18 &  -24:27:04.8 &$   207.5 \pm 0.3 $&   5 &  17.8 \\ 
103 &  1540 &  05:24:13.19 &  -24:33:22.3 &$   204.0 \pm 0.2 $&   5 &  17.9 \\ 
104 &  1371 &  05:24:13.90 &  -24:32:34.5 &$   202.8 \pm 0.2 $&   5 &  17.7 \\ 
105 &   968 &  05:24:14.11 &  -24:31:49.6 &$   202.1 \pm 0.2 $&   5 &  17.0 \\ 
106 &  1199 &  05:24:14.11 &  -24:32:38.7 &$   211.9 \pm 0.1 $&   5 &  17.6 \\ 
107 &  1683 &  05:24:14.18 &  -24:30:43.8 &$   207.7 \pm 0.2 $&   5 &  18.1 \\ 
108 &  1263 &  05:24:14.20 &  -24:32:47.9 &$   204.6 \pm 0.2 $&   5 &  17.7 \\ 
109 &  2009 &  05:24:14.26 &  -24:27:19.7 &$   207.4 \pm 0.4 $&   5 &  18.6 \\ 
110 &  1971 &  05:24:14.99 &  -24:32:15.6 &$   203.1 \pm 0.4 $&   5 &  18.5 \\ 
111 &  1442 &  05:24:15.06 &  -24:30:01.7 &$   210.3 \pm 0.4 $&   5 &  17.9 \\ 
112 &  1888 &  05:24:15.09 &  -24:32:53.8 &$   197.9 \pm 0.2 $&   5 &  18.4 \\ 
113 &  1577 &  05:24:15.28 &  -24:31:56.6 &$   207.2 \pm 0.2 $&   5 &  18.0 \\ 
114 &  1250 &  05:24:15.41 &  -24:32:35.2 &$   206.8 \pm 0.1 $&   5 &  17.6 \\ 
115 &  2704 &  05:24:15.63 &  -24:28:37.6 &$   202.7 \pm 0.3 $&   4 &  19.0 \\ 
116 &  1834 &  05:24:15.78 &  -24:30:37.7 &$   206.7 \pm 0.4 $&   5 &  18.4 \\ 
117 &  1555 &  05:24:15.81 &  -24:30:52.6 &$   205.0 \pm 0.2 $&   5 &  18.1 \\ 
118 &  1427 &  05:24:15.94 &  -24:32:20.2 &$   200.9 \pm 0.1 $&   5 &  17.9 \\ 
119 &  1060 &  05:24:16.03 &  -24:30:37.5 &$   211.6 \pm 0.1 $&   5 &  17.2 \\ 
120 &  1341 &  05:24:16.35 &  -24:29:37.3 &$   202.2 \pm 0.2 $&   5 &  17.7 \\ 
121 &  1802 &  05:24:16.57 &  -24:35:55.5 &$   204.7 \pm 0.3 $&   5 &  18.2 \\ 
122 &  2400 &  05:24:16.70 &  -24:32:17.1 &$   204.3 \pm 0.7 $&   5 &  19.1 \\ 
123 &  1932 &  05:24:16.71 &  -24:33:00.5 &$   209.5 \pm 0.3 $&   5 &  18.4 \\ 
124 &  1522 &  05:24:17.09 &  -24:27:57.3 &$   206.8 \pm 0.4 $&   5 &  18.0 \\ 
125 &  1715 &  05:24:17.15 &  -24:32:29.4 &$   210.4 \pm 0.3 $&   5 &  18.2 \\ 
126 &  2150 &  05:24:17.33 &  -24:28:50.7 &$   204.5 \pm 0.6 $&   5 &  18.6 \\ 
127 &  1212 &  05:24:17.37 &  -24:30:54.2 &$   202.3 \pm 0.2 $&   5 &  17.5 \\ 
128 &  1208 &  05:24:17.42 &  -24:29:38.5 &$   208.1 \pm 0.1 $&   5 &  17.5 \\ 
129 &  1909 &  05:24:17.74 &  -24:31:09.1 &$   209.5 \pm 0.7 $&   5 &  18.4 \\ 
130 &  1596 &  05:24:17.83 &  -24:32:08.1 &$   209.8 \pm 0.3 $&   5 &  18.0 \\ 
131 &  1498 &  05:24:17.98 &  -24:38:23.9 &$   204.3 \pm 0.2 $&   5 &  17.9 \\ 
132 &  1058 &  05:24:18.60 &  -24:24:55.8 &$   202.5 \pm 0.2 $&   5 &  17.2 \\ 
133 &  2551 &  05:24:18.74 &  -24:28:30.8 &$   205.1 \pm 1.4 $&   5 &  19.0 \\ 
134 &  1762 &  05:24:19.31 &  -24:27:53.6 &$   210.4 \pm 0.5 $&   5 &  18.3 \\ 
135 &  1145 &  05:24:19.67 &  -24:31:06.3 &$   209.1 \pm 0.1 $&   5 &  17.4 \\ 
136 &  1597 &  05:24:19.93 &  -24:32:46.0 &$   203.3 \pm 0.4 $&   5 &  18.1 \\ 
137 &  1304 &  05:24:20.17 &  -24:34:08.4 &$   205.5 \pm 0.1 $&   5 &  17.6 \\ 
138 &  1015 &  05:24:20.33 &  -24:33:14.7 &$   202.6 \pm 0.1 $&  10 &  17.5 \\ 
139 &  2539 &  05:24:20.36 &  -24:25:20.6 &$    35.4 \pm 0.2 $&   5 &  18.3 \\ 
140 &  1728 &  05:24:20.41 &  -24:26:40.8 &$   204.9 \pm 0.3 $&   5 &  18.2 \\ 
141 &  1820 &  05:24:20.49 &  -24:28:54.4 &$   203.2 \pm 0.3 $&   5 &  18.3 \\ 
142 &  2336 &  05:24:21.36 &  -24:29:37.3 &$   207.0 \pm 0.5 $&   5 &  18.8 \\ 
143 &  2394 &  05:24:21.57 &  -24:29:12.4 &$   208.1 \pm 0.8 $&   5 &  18.8 \\ 
144 &  2393 &  05:24:21.60 &  -24:32:16.2 &$   206.6 \pm 0.6 $&   5 &  18.7 \\ 
145 &  1422 &  05:24:22.04 &  -24:31:01.3 &$   208.8 \pm 0.4 $&   5 &  17.8 \\ 
146 &  1069 &  05:24:22.39 &  -24:31:51.3 &$   204.9 \pm 0.1 $&   5 &  17.2 \\ 
147 &  2703 &  05:24:22.68 &  -24:30:55.1 &$   209.9 \pm 0.6 $&   5 &  19.0 \\ 
148 &  1810 &  05:24:22.81 &  -24:33:43.7 &$   211.4 \pm 0.2 $&   5 &  18.2 \\ 
149 &  1858 &  05:24:23.14 &  -24:31:30.7 &$   204.1 \pm 0.7 $&   5 &  18.3 \\ 
150 &  1123 &  05:24:23.16 &  -24:25:56.0 &$   205.6 \pm 0.3 $&   5 &  17.4 \\ 
151 &  1804 &  05:24:23.55 &  -24:34:18.9 &$   208.1 \pm 0.2 $&   5 &  18.2 \\ 
152 &  2574 &  05:24:23.78 &  -24:25:49.1 &$   208.6 \pm 1.1 $&   5 &  19.0 \\ 
153 &  2048 &  05:24:24.04 &  -24:32:13.5 &$   205.5 \pm 0.4 $&   5 &  18.5 \\ 
154 &  1749 &  05:24:24.68 &  -24:31:40.4 &$   201.9 \pm 0.2 $&   5 &  18.2 \\ 
155 &  1890 &  05:24:25.61 &  -24:35:15.9 &$   207.0 \pm 0.3 $&   5 &  18.3 \\ 
156 &  2606 &  05:24:25.79 &  -24:30:39.1 &$   202.2 \pm 1.1 $&   5 &  19.1 \\ 
157 &  2033 &  05:24:26.47 &  -24:23:14.5 &$    11.1 \pm 0.2 $&   5 &  18.1 \\ 
158 &  2345 &  05:24:26.75 &  -24:32:23.0 &$   207.8 \pm 0.7 $&   5 &  18.7 \\ 
159 &  1973 &  05:24:27.93 &  -24:29:26.5 &$   209.7 \pm 0.5 $&   5 &  18.5 \\ 
160 &  1373 &  05:24:28.38 &  -24:25:16.0 &$    38.1 \pm 0.3 $&   5 &  17.2 \\ 
\hline
\hline
\end{tabular}
\end{table}
\addtocounter{table}{-1}

\begin{table}[b]
\scriptsize
\caption{Radial velocities of stars in NGC 1904 -- continued}
\begin{tabular}{cccccccc}
\# & ID &   RA   &  DEC   & Vel   &   N     & V mag\\
   &    & (2000) & (2000) & [km/s]    & & \\
\hline
161 &  1410 &  05:24:28.65 &  -24:31:43.6 &$   204.3 \pm 0.3 $&   5 &  17.9 \\ 
162 &  2597 &  05:24:28.69 &  -24:28:05.9 &$   208.0 \pm 0.9 $&   5 &  19.0 \\ 
163 &  2354 &  05:24:28.78 &  -24:22:57.7 &$    76.9 \pm 1.7 $&   4 &  18.3 \\ 
164 &  1939 &  05:24:29.02 &  -24:37:35.7 &$   206.0 \pm 0.5 $&   5 &  18.4 \\ 
165 &  2434 &  05:24:29.18 &  -24:30:42.4 &$   204.5 \pm 0.4 $&   5 &  18.7 \\ 
166 &  2367 &  05:24:29.51 &  -24:27:36.1 &$   202.5 \pm 1.1 $&   5 &  18.8 \\ 
167 &  1151 &  05:24:30.97 &  -24:31:41.4 &$   -21.5 \pm 0.1 $&   5 &  17.2 \\ 
168 &  2843 &  05:24:31.06 &  -24:30:45.9 &$   207.2 \pm 1.3 $&   3 &  19.1 \\ 
169 &  1439 &  05:24:38.78 &  -24:37:15.9 &$    40.5 \pm 0.2 $&   5 &  17.9 \\ 
170 &  2115 &  05:24:41.18 &  -24:33:52.3 &$   209.1 \pm 0.7 $&   5 &  18.6 \\ 
171 &  1190 &  05:24:41.26 &  -24:33:41.2 &$   207.6 \pm 0.2 $&   5 &  17.5 \\ 
172 &  1702 &  05:24:46.57 &  -24:34:31.5 &$   205.4 \pm 0.6 $&   4 &  18.2 \\ 
173 &  1162 &  05:24:53.79 &  -24:36:01.8 &$    80.4 \pm 0.1 $&  10 &  17.5 \\ 
\hline
\hline
\end{tabular}
\end{table}

\begin{table}[b]
\caption{Radial velocity dispersion for NGC 1851}
\begin{tabular}{ccccc}
Bin [pc]   & N stars & Bin center & $\sigma$ [km/s]\\
\hline
$    0 -    5 $&$  20  $&$  4.05 \pm  0.77  $&$  7.24 \pm  1.16 $\\ 
$    5 -    8 $&$  27  $&$  6.64 \pm  0.84  $&$  5.78 \pm  0.80 $\\ 
$    8 -   11 $&$  43  $&$  9.55 \pm  0.91  $&$  4.92 \pm  0.54 $\\ 
$   11 -   14 $&$  29  $&$ 12.40 \pm  0.85  $&$  4.10 \pm  0.56 $\\ 
$   14 -   17 $&$  20  $&$ 15.05 \pm  0.93  $&$  4.40 \pm  0.72 $\\ 
$   17 -   24 $&$  24  $&$ 20.14 \pm  1.87  $&$  4.52 \pm  0.67 $\\ 
$   24 -   40 $&$  21  $&$ 29.30 \pm  4.12  $&$  3.38 \pm  0.55 $\\ 
\hline
\hline
\end{tabular}
\label{tabDisp1851}
\end{table}

\begin{table}[b]
\caption{Radial velocity dispersion for NGC 1904}
\begin{tabular}{ccccc}
Bin [pc]   & N stars & Bin center & $\sigma$ [km/s]\\
\hline
$    0 -    4 $&$  14  $&$  3.20 \pm  0.73  $&$  4.01 \pm  0.78 $\\ 
$    4 -    7 $&$  37  $&$  5.66 \pm  0.83  $&$  3.78 \pm  0.46 $\\ 
$    7 -   10 $&$  20  $&$  8.65 \pm  0.95  $&$  3.34 \pm  0.55 $\\ 
$   10 -   13 $&$  24  $&$ 11.56 \pm  1.02  $&$  2.53 \pm  0.40 $\\ 
$   13 -   16 $&$  19  $&$ 14.47 \pm  0.97  $&$  2.73 \pm  0.48 $\\ 
$   16 -   21 $&$  17  $&$ 17.94 \pm  1.49  $&$  1.75 \pm  0.35 $\\ 
$   21 -   35 $&$  15  $&$ 27.09 \pm  3.74  $&$  2.28 \pm  0.46 $\\ 
\hline
\hline
\end{tabular}
\label{tabDisp1904}
\end{table}

\begin{table*}[b]
\label{tabSummary}
\caption{Clusters with velocity dispersion data probing accelerations below $a_0$.}
\begin{tabular}{lccccccccc}
     name            & R$_{\odot}$ & R$_{MW}$& M(v)   &    Mass        &  $r_0$   & $r_{tidal}$&    $r_{flat}$    &  $\sigma$     & a @ $r_{flat}$  \\
                       &  [kpc] & [kpc]  &          & [M$_\odot$]     & [pc]     &  [pc]   &        [pc]       &     [km/s]    & [$10^{-8}$ cm s$^{-2}$] \\
\hline
NGC 1851               &$ 12.1 $&$ 16.7 $&$ -8.33  $&$ 1.8\times10^{5} $&$  14.5  $&$  41   $&$  12.5 \pm 2.5 $&$ 4.0 \pm 0.5 $&$ 1.6 \pm 0.4$ \\
NGC 1904 (M79)         &$ 12.9 $&$ 18.8 $&$ -7.86  $&$ 1.2\times10^{5} $&$  11.7  $&$  31   $&$    12 \pm 2   $&$ 2.25\pm 0.4 $&$ 1.1 \pm 0.3$ \\
NGC 5139 ($\omega$ Cen)&$  5.5 $&$  6.4 $&$ -10.29 $&$ 1.1\times10^{6} $&$  35.7  $&$  72   $&$    32 \pm 3   $&$   7\pm 1    $&$ 1.5 \pm 0.4$ \\
NGC 6171 (M107)        &$  6.4 $&$  3.3 $&$ -7.13  $&$ 5.9\times10^{4} $&$   8.0  $&$  32   $&$     8 \pm 2   $&$ 2.7\pm 0.3  $&$ 1.3 \pm 0.6$ \\
NGC 6341 (M92)         &$  8.2 $&$  9.6 $&$ -8.20  $&$ 1.6\times10^{5} $&$  13.6  $&$  36   $&$    12 \pm 2   $&$ 3.1\pm 0.4  $&$ 1.5 \pm 0.6$ \\
NGC 7078 (M15)         &$ 10.3 $&$ 10.4 $&$ -9.17  $&$ 3.9\times10^{5} $&$  21.3  $&$  64   $&$    20 \pm 2   $&$ 3.2\pm 0.5  $&$ 1.4 \pm 0.4$ \\
NGC 7099 (M30)         &$  8.0 $&$  7.1 $&$ -7.43  $&$ 7.8\times10^{4} $&$   9.6  $&$  43   $&$    10 \pm 2   $&$ 2.2\pm 0.3  $&$ 1.1 \pm 0.4$ \\
\hline
\hline
\end{tabular}
\\Note to table: The meaning of the columns is as follow. 1) cluster
name. 2) distance of the clusters from the sun. 3) Distance from Milky
Way center.  4) Absolute total V band magnitude, used to derive the
mass assuming M/L=1 in solar units. 5) Cluster mass from luminosity in
solar masses.  6) Radius where the acceleration is $a_0=1.2\times
10^{-8}$ cm s$^{-2}$. 7) Cluster tidal radius from \cite{harris96}. 8)
Radius where the velocity dispersion profile flattens out.  9) Cluster
asymptotic velocity dispersion. 10) Acceleration at the radius
$r_{flat}$.
\end{table*}

\end{document}